\documentclass[acmsmall,10pt,screen]{acmart}

\AtBeginDocument{%
  }
    
\usepackage{algorithmic}
\usepackage{textcomp}
\usepackage{color}
 \usepackage{booktabs}

\usepackage[linesnumbered,ruled,vlined]{algorithm2e}
\SetKwInput{KwInput}{Input}                
\SetKwInput{KwOutput}{Output}              

\usepackage{float}
\usepackage{caption}
\usepackage{subcaption}
\usepackage{xspace}
\usepackage{enumitem}
\usepackage{hyperref}

\newcommand\tool{\textit{GUIPilot}\xspace}

\newcommand\screen{\mathbf{s}\xspace}
\newcommand\screenset{\mathcal{S}\xspace}
\newcommand\widget{\mathbf{w}\xspace}
\newcommand\widgetset{\mathcal{W}\xspace}
\newcommand\action{\mathbf{a}\xspace}
\newcommand\actionchain{\mathbf{ac}\xspace}
\newcommand\actionchainset{\mathcal{AC}\xspace}
\newcommand\process{\mathbf{p}\xspace}

\usepackage{diagbox}

\usepackage{tcolorbox}
\usepackage[framemethod=TikZ]{mdframed}

\usepackage{varwidth}
\usepackage{tcolorbox}
\newcommand{\coloredbox}[2]{%
    \colorbox{#1}{\begin{varwidth}[t]{\dimexpr\linewidth-2\fboxsep\relax}#2\end{varwidth}}%
}

\definecolor{SoftRed}{RGB}{255, 228, 225}
\definecolor{SoftGreen}{RGB}{230, 255, 230}
\definecolor{SoftOrange}{RGB}{255, 229, 204}
\definecolor{SoftPurple}{RGB}{224, 224, 248}
\definecolor{SoftBlue}{RGB}{202, 240, 255}
\definecolor{SoftGray}{RGB}{237, 237, 237}
\definecolor{SoftYellow}{RGB}{255, 255, 204}

\setcopyright{cc}
\setcctype{by}
\acmDOI{10.1145/3728909}
\acmYear{2025}
\acmJournal{PACMSE}
\acmVolume{2}
\acmNumber{ISSTA}
\acmArticle{ISSTA034}
\acmMonth{7}

\begin{document}

\title{GUIPilot: A Consistency-Based Mobile GUI Testing Approach for Detecting Application-Specific Bugs}

\author{Ruofan Liu}
\authornote{Both authors contributed equally to this research.}
\orcid{0009-0002-9440-3152}
\affiliation{%
  \institution{Shanghai Jiao Tong University}
  \city{Shanghai}
  \country{China}
}
\affiliation{%
  \institution{National University of Singapore}
  \city{Singapore}
  \country{Singapore}
}
\email{liu.ruofan16@u.nus.edu}

\author{Xiwen Teoh}
\authornotemark[1]
\orcid{0009-0009-8528-9088}
\affiliation{%
  \institution{Shanghai Jiao Tong University}
  \city{Shanghai}
  \country{China}
}
\affiliation{%
  \institution{National University of Singapore}
  \city{Singapore}
  \country{Singapore}
}
\email{xiwen.teoh@u.nus.edu}

\author{Yun Lin}
\authornote{Corresponding author.}
\orcid{0000-0001-8255-0118}
\affiliation{%
  \institution{Shanghai Jiao Tong University}
  \city{Shanghai}
  \country{China}
}
\email{lin_yun@sjtu.edu.cn}

\author{Guanjie Chen}
\orcid{0009-0008-6846-131X}
\affiliation{%
  \institution{Shanghai Jiao Tong University}
  \city{Shanghai}
  \country{China}
}
\email{guanjie.chen@sjtu.edu.cn}

\author{Ruofei Ren}
\orcid{0009-0000-5189-2860}
\affiliation{%
  \institution{Shanghai Jiao Tong University}
  \city{Shanghai}
  \country{China}
}
\email{renruofei0120@sjtu.edu.cn}

\author{Denys Poshyvanyk}
\orcid{0000-0002-5626-7586}
\affiliation{%
  \institution{College of William and Mary}
  \city{Williamsburg}
  \country{USA}
}
\email{denys@cs.wm.edu}

\author{Jin Song Dong}
\orcid{0000-0002-6512-8326}
\affiliation{%
  \institution{National University of Singapore}
  \city{Singapore}
  \country{Singapore}
}
\email{dcsdjs@nus.edu.sg}

\begin{abstract}
GUI testing is crucial for ensuring the reliability of mobile applications. 
State-of-the-art GUI testing approaches are successful in 
exploring more application scenarios and
discovering \textit{general} bugs such as application crashes.
However, industrial GUI testing also needs to 
investigate \textit{application-specific} bugs such as deviations in screen layout, widget position, 
or GUI transition from the GUI design mock-ups created by the application designers.
These mock-ups specify the expected screens, widgets, and their respective behaviors.
Validating the consistency between the GUI design and the implementation
is labor-intensive and time-consuming, 
yet, this validation step plays an important role in industrial GUI testing.

In this work, we propose \tool, an approach for detecting inconsistencies between the mobile design and their implementations.
The mobile design usually consists of design mock-ups that specify 
(1) the expected screen appearances (e.g., widget layouts, colors, and shapes) and 
(2) the expected screen behaviors, regarding how one screen can transition into another (e.g., labeled widgets with textual description).
Given a design mock-up and the implementation of its application,
\tool reports both their screen inconsistencies as well as process inconsistencies. 
On the one hand, \tool detects the screen inconsistencies 
by abstracting every screen into a widget container where each widget
is represented by its position, width, height, and type.
By defining the partial order of widgets and the costs of replacing, inserting, and deleting widgets in a screen,
we convert the screen-matching problem into an optimizable widget alignment problem.
On the other hand, 
we translate the specified GUI transition into 
stepwise actions on the mobile screen (e.g., click, long-press, input text on some widgets).
To this end, we propose a \textit{visual prompt} 
for the vision-language model to infer widget-specific actions on the screen.
By this means, we can validate the presence or absence of expected transitions in the implementation.
Our extensive experiments on 80 mobile applications
and 160 design mock-ups show that
(1) \tool can achieve 99.8\% precision and 98.6\% recall in detecting screen inconsistencies,
outperforming the state-of-the-art approach, such as GVT, by 66.2\% and 56.6\% respectively, and
(2) \tool reports zero errors in detecting process inconsistencies.
Furthermore, our industrial case study on applying \tool on a trading mobile application shows that
\tool has detected nine application bugs, 
and all the bugs were confirmed by the original application experts.
Our code is available at \url{https://github.com/code-philia/GUIPilot}. 

%

\end{abstract}



\received{2024-10-31}
\received[accepted]{2025-03-31}

\maketitle

\section{Introduction}



\begin{figure}[t]
    \centering
    \includegraphics[width=\linewidth]{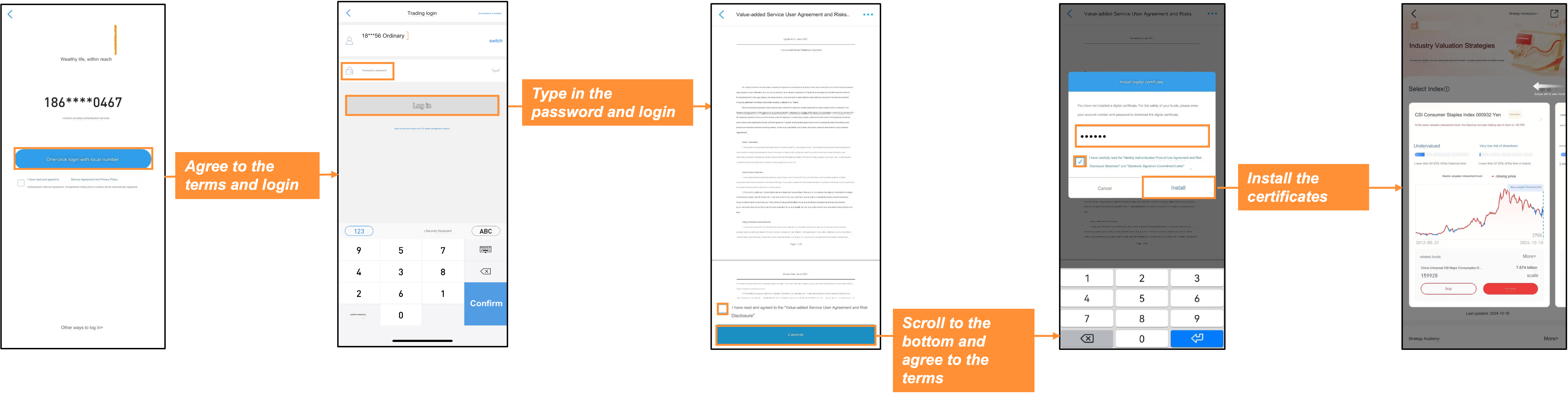}
    \caption{A design mock-up on the login process of a trading mobile application, 
    consisting of five screens and four transitions.}
    \label{fig:mockup-example}
\end{figure}


Recent decades have witnessed the global mobile application market grow to \$228.98 billion in 2023,
and the market is projected to expand at an annual growth rate of 14.3\% from 2024 to 2030 \cite{mobile-app-market}.
Rigorous GUI testing is crucial to ensure the reliability of critical mobile applications in trading, banking, and government services \cite{gui-testing-market}.
A typical industrial application development life-cycle includes the following stages \cite{gui-development-cycle}:
\begin{itemize}[leftmargin=*]
  \item \textbf{Design Stage (design mock-ups creation):}
    87\% of application designers utilize prototyping tools to streamline their design workflow \cite{mockup-commonality}. 
   They generate high-fidelity \textit{mock-ups} using popular prototyping tools such as Sketch \cite{sketch-app}, Axure \cite{axure-app}, and Balsamiq \cite{balsamiq-app}. 
    These mock-ups
    (1) visualize interface layouts and designs, detailing the appearance of widgets, buttons, icons, typography, and
    (2) illustrate how one screen can transition into another by textual description, serving as the \textit{application specification}.
    \autoref{fig:mockup-example} shows an example of partial design mock-ups.
    This design mock-up demonstrates the expected login process, 
    showing how the screens are interconnected through various interactive widgets.
  \item \textbf{Development Stage:}
    Based on the specifications indicated by the mock-ups,
    developers implement the GUI and the underlying functionalities of the widgets.
  \item \textbf{Testing Stage (design mock-ups validation):}
    Finally, application testers validate the consistency between the design mock-ups and the implementation,
    by writing GUI test cases and scripts.
    The discovered inconsistencies are reported as bugs to be fixed.
\end{itemize}

Testing mobile applications against the design mock-ups is non-trivial.
First, a screen can consist of dozens of widgets and
a manual comparison of two screens (one from the application and the other from the design mock-ups)
can be fairly error-prone.
Second, although the design mock-ups provide textual descriptions,
the screen transition requires sufficient domain knowledge to complete all the triggering actions.
For example, in the fourth screen of \autoref{fig:mockup-example}, one must enter a password and agree to the terms by clicking the appropriate checkbox before the ``Install'' button becomes functional.
Failing to complete these steps cannot initiate the operability of the ``Install'' button.
Third, the mobile application can be updated frequently,
the updates in the application may require a user to go through screen comparison and transition validation over and over again.


\begin{figure}[h]
    \centering
    \includegraphics[width=0.5\linewidth]{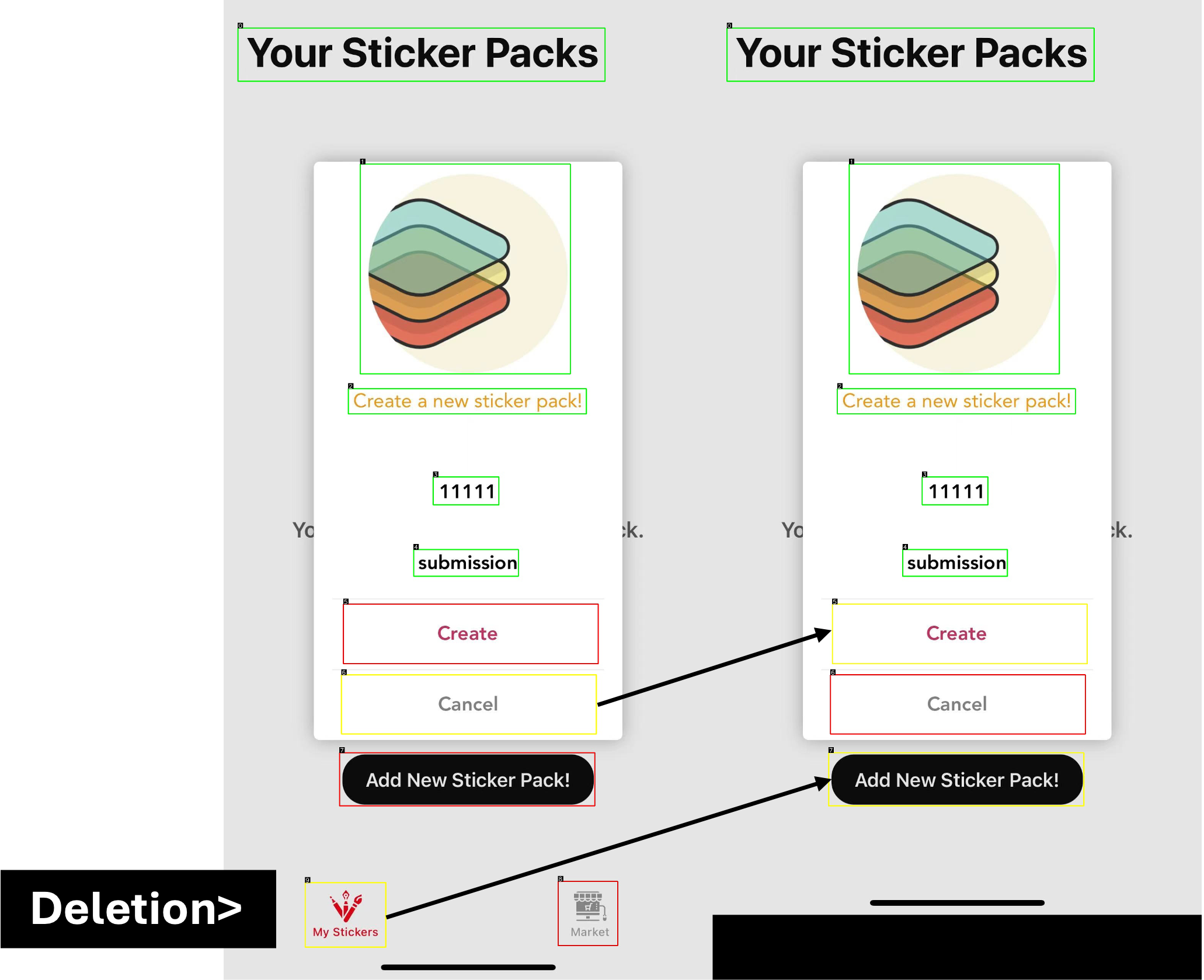}
    \caption{A failure case of GVT.
    \coloredbox{SoftRed}{Red} boxes are widgets that could not be matched.
    \textbf{Lines} highlight the matches.}
    \label{fig:gvt-issues}
\end{figure}

Existing GUI testing solutions are generally designed to
explore more unseen scenarios and discover general bugs such as
application crashes \cite{7515457,liu2024testing, huang2024crashtranslator, wang2022detecting, su2021fully, zhao2022recdroid+, zhao2020seenomaly, liu2022nighthawk, su2017guided}.
While those approaches are effective in detecting general bugs,
there is little work on testing the application
against the design mock-ups.
The most relevant work is a GVT technique by Moran et al. \cite{moran2018automated},
which detects the inconsistency
between a screen in an application and a mocking screen.
However, their work still suffers from the following challenges:
\begin{itemize}[leftmargin=*]
  \item \textbf{Lack of Transition Automation}:
    The GVT approach does not consider testing the screen transition,
    which is a crucial part of the specification in the design mock-ups.
  \item \textbf{Accurate Screen Matching}:
    Technically, GVT matches the widgets by \textit{relative} position in the screen,
    which can miss important layout semantics.
    \autoref{fig:gvt-issues} illustrates an example where mismatches occur when the last row of widgets is deleted.
    As a result, the `Cancel' button is mismatched to the `Create' button, 
    and the `My stickers' icon is mismatched to the `Add new sticker pack' button 
    because the matched widgets share similar relative positions on the screen.
    This leads to both false positives and false negatives.
\end{itemize}



In this work, we propose \tool, an approach for detecting inconsistencies between the mobile design and its implementation,
reporting both screen and process (i.e., transition) inconsistencies.

To detect screen inconsistency,
\tool abstracts every screen into a widget container where each widget
is represented by its position, width, height, and type.
Then, we convert the screen matching problem to an optimizable widget alignment problem
by defining the partial order of widgets and the costs of replacing, inserting, and deleting widgets in a screen.
By this means, we can mitigate the local matching issues in GVT and
compare two screens regarding their global layout semantics.

To detect process inconsistency,
we translate the GUI transition specified in the mock-up design into
stepwise actions on the mobile screen (e.g., click, long-press, input text on some widgets)
based on the state-of-the-art vision-language model (VLM).
To mitigate the potential hallucination of VLM,
we propose the \textit{visual prompt} technique for VLM,
forcing the model to infer actions only from the relevant widgets with limited action options.
By this means, we can navigate to the next screen according to the design mock-ups and validate whether an expected transition can happen or be missed in the implementation.

We evaluate \tool on 80 mobile applications
with mock-up designs covering four application types and 160 design mock-ups,
showing that
(1) \tool can achieve the precision of 94.5\% and the recall of 99.6\%
in detecting screen inconsistencies,
outperforming the state-of-the-art approaches, such as GVT, by 66.2\% and 56.6\% respectively, 
(2) \tool reports zero errors in detecting process inconsistencies, and
(3) \tool is efficient in that the screen matching algorithm takes on average 0.001s and
the transition takes an average of 0.19s.
Further, we conduct a case study on applying \tool on a trading application with 32 million users\footnote{For the sake of anonymity, we do not reveal the name of the mobile application in the submission} with our industrial collaborator.
The results demonstrate that
\tool detects nine application inconsistency bugs,
and all the bugs have been confirmed by the application experts.


In summary, the contributions of this work are as follows:
\begin{itemize}[leftmargin=*]
    \item We propose \tool, a solution to systematically detect inconsistencies between the design mock-ups and the mobile application implementation.
    To the best of our knowledge, \tool is the first end-to-end GUI testing solution tailored for design mock-ups that are widely adopted in the industry.
    
    \item We technically address the screen matching problem regarding global layout semantics for screen consistency and the design-to-action problem for process consistency.
    The experiments show that we address both problems with high accuracy.
    \item We deliver the \tool as a web application\footnote{We have released an anonymous code repository at \url{https://anonymous.4open.science/r/guipilot-C65C}.},
    enabling the research community and industry to conduct further research and applications.
    
    \item We conducted experiments on diverse types of mobile applications,
        showing the effectiveness of \tool.
        Further, we show that \tool is able to detect real design violations on an industrial trading application.
\end{itemize}

Given the space limitations,
more tool demos, videos, and experimental details are available at \cite{guipilot-website}.

\section{Problem Statement}\label{sec:ps}

In this section, we provide the formal definitions and problem statements.

\noindent\textbf{Screen.}
We consider a screen $\screen$ as a single GUI screen,
containing multiple widgets including images, buttons, text, etc.
Each widget $\widget$ is represented in $\langle x, y, w, h, t \rangle$ format.
Specifically, $x, y$ are the top-left corner coordinates of the widget on the screen, using the screenshot's top-left corner as the origin.
$w, h$ are the width and height of the widget, relative to the full screenshot size.
And $t$ is the type of the widget.
We define 7 major types of widgets as detailed in Table \ref{tab:widget-class}.
A screen contains a collection of widgets, i.e. $\widgetset = \{\widget_i|\widget_i=\langle x, y, w, h, t \rangle\}$.

\begin{table}[h!]
    \centering
    \small
    \caption{Widget categories}\label{tab:widget-class}
    \resizebox{0.7\textwidth}{!}{%
    \begin{tabular}{ll}
        \toprule
        \textbf{Interactable} & \textbf{Non-Interactable} \\ \midrule
        TextButton  \includegraphics[width=1cm, height=0.4cm]{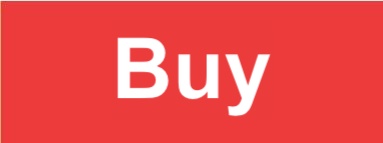}\par\includegraphics[width=1cm, height=0.4cm]{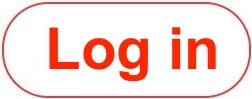}
        &
        TextView  \includegraphics[width=1cm, height=0.4cm]{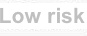}\par\includegraphics[width=1.2cm, height=0.4cm]{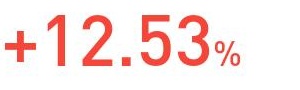} \\ \midrule

        IconButton  \fboxsep=0pt\fbox{\includegraphics[width=0.5cm, height=0.5cm]{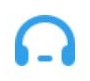}}\par\fboxsep=0pt\fbox{\includegraphics[width=0.5cm, height=0.5cm]{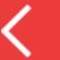}}\par\fboxsep=0pt\fbox{\includegraphics[width=0.5cm, height=0.5cm]{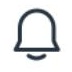}}\par\includegraphics[width=1cm, height=0.5cm]{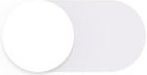}
        &
        ImageView \fboxsep=0pt\fbox{\includegraphics[width=0.5cm, height=0.5cm]{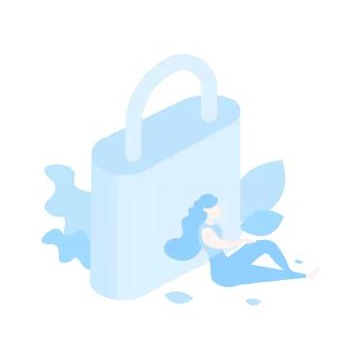}}\par\fboxsep=0pt\fbox{\includegraphics[width=0.5cm, height=0.5cm]{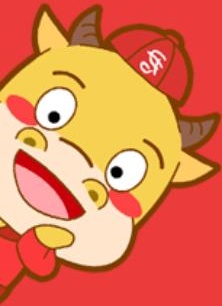}}\par\fboxsep=0pt\fbox{\includegraphics[width=0.5cm, height=0.5cm]{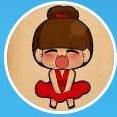}}  \\ \midrule

        CombinedButton \fboxsep=0pt\fbox{\includegraphics[width=1.2cm, height=1cm]{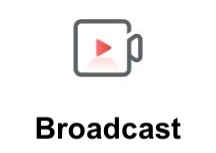}}\par\fboxsep=0pt\fbox{\includegraphics[width=1.2cm, height=1cm]{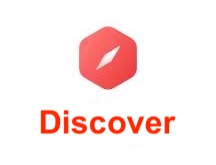}}
        &
        Chart  \fboxsep=0pt\fbox{\includegraphics[width=3cm, height=1cm]{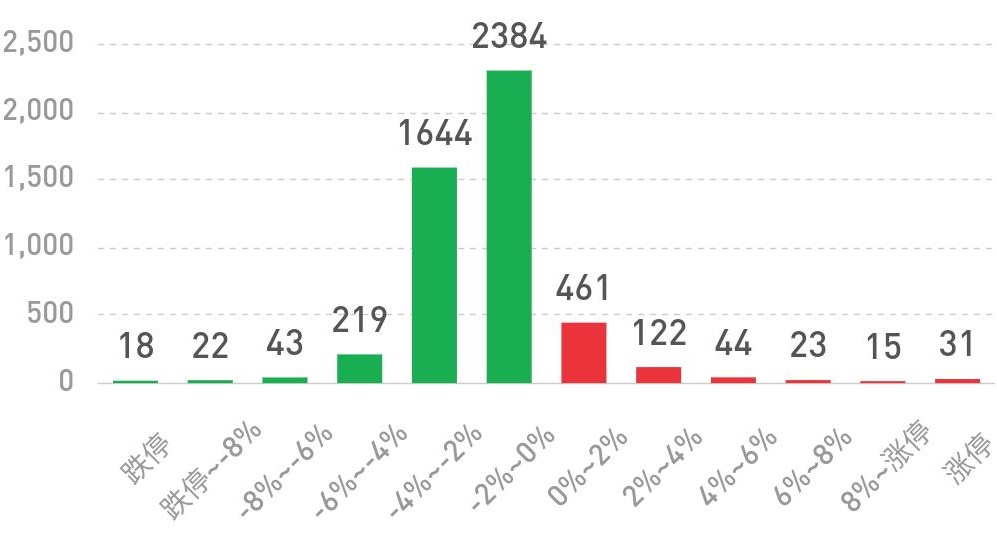}}  \\ \midrule

        \multicolumn{2}{l}{InputBox \fboxsep=0pt\fbox{\includegraphics[width=6cm, height=0.6cm]{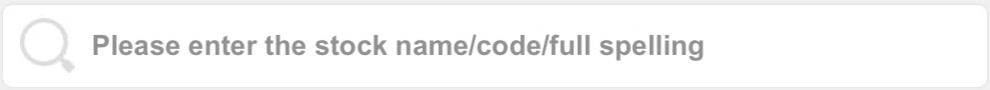}}}   \\
        \bottomrule
    \end{tabular}
    }
\end{table} 

As in \autoref{tab:widget-class}, we classify widgets into interactable and non-interactable.
Interactable widgets include text buttons, icon buttons, combined buttons, and input boxes.
A text button conveys its function through text, whereas an icon button uses an icon.
A combined button integrates both text and an icon to provide a more comprehensive representation.
An input box allows users to enter text.
Non-interactable widgets include text views, image views, and charts.

\noindent\textbf{Action.}
On a single screen $\screen$, one can perform a set of actions, each interacting with different widgets.
We define the action space to include 7 actions: click, long press, send keys, scroll, swipe, drag and drop, and go back.
An action chain $\actionchain$ is defined as a sequence of actions $\actionchain = (\action_1(\widget_1), \action_2(\widget_2), ..., \action_i(\widget_i))$.
Taking \autoref{fig:mockup-example} as an example, the action chain from screen 4 to screen 5 is $\actionchain$ = (click(the password input box), send\_keys(our password), click(the agreement checkbox), click(the install button)).

\noindent\textbf{Process.}
A process $\process$ is a connected directed graph $\process: G = ( \screenset, \actionchainset )$,
where $\screenset$ are a set of screens, and $\actionchainset$ are a set of action chains.
Each action chain in $\actionchain \in \actionchainset$ can either lead to a transition from one screen to another or result in staying on the same screen, specifying the edges in the graph.
For example, \autoref{fig:mockup-example} consists of 5 vertices (5 screens), and they are connected by 4 edges (4 action chains or transitions).


\noindent\textbf{Screen Inconsistency.}
We consider three types of inconsistencies as screen inconsistencies, i.e.,
extra widgets, missing widgets, and semantic change.
Formally, we denote the set of widgets on the mock-up for a specific screen as $\widgetset^{\textit{tar}}$,
and the set of widgets in the implementation for the corresponding screen as $\widgetset$.
If there exists a ground-truth matching function $f(.)$ that takes two sets of widgets and returns the sets of matched pairs, i.e. $f(\widgetset^{\textit{tar}}, \widgetset) = \{(\widget^{\textit{tar}}_i, \widget_j) | \widget^{\textit{tar}}_i \in \widgetset^{\textit{tar}}, \widget_j \in \widgetset\, \textit{and they match}\}$.
The inconsistency is reported when any of the conditions are met:

\textit{(i) Missing widget(s) in the implementation.}
\begin{equation}\label{eq:miss-widget}
    \exists \widget^{\textit{tar}}_i \in \widgetset^{\textit{tar}} \textit{ such that } \widget^{\textit{tar}}_i \notin f(\widgetset^{\textit{tar}}, \widgetset)
\end{equation}

\textit{(ii) Extra widget(s) in the implementation.}
\begin{equation}\label{eq:extra-widget}
    \exists \widget_j \in \widgetset \textit{ such that } \widget_j \notin f(\widgetset^{\textit{tar}}, \widgetset)
\end{equation}

\textit{(iii) Semantic change in matched widgets}:
Two widgets are successfully paired but exhibit different semantics.
For instance, their text, color, or widget type may have been altered.
Let $\epsilon_s$ be the threshold for acceptable semantic changes, and let $g(.)$ represent a semantic extraction function that maps a widget to a d-dimensional representation vector.
The semantic difference between two paired widgets can then be quantified as the distance between their semantic vectors:
\begin{equation}\label{eq:semantic-change}
    \begin{aligned}
    \exists & (\widget^{\textit{tar}}_i, \widget_j) \in f(\widgetset^{\textit{tar}}, \widgetset) \textit{ such that } \\
    &  t_i \neq t_j \textit{ or } \| g(\widget^{\textit{tar}}_i) - g(\widget_j) \|_2 > \epsilon_s\\
    \end{aligned}
\end{equation}


\noindent\textbf{Process Inconsistency.}
Process inconsistency detection is theoretically a graph comparison problem
considering topological isomorphism. 
Let the expected process on the mock-up be denoted as $\process^{\textit{tar}}: G = ( \screenset^{\textit{tar}}, \actionchainset^{\textit{tar}} )$.
$\actionchainset^{\textit{tar}}$ may contain ambiguous instructions, necessitating the auto-completion of the missing instructions.
This results in the corrected action chain, $\actionchainset^{\textit{tar}}_{\textit{complete}}$.
Executing $\actionchainset^{\textit{tar}}_{\textit{complete}}$ on the implementation generates a process graph on the app $\process: G = (\screenset, \actionchainset^{\textit{tar}}_{\textit{complete}})$.
By comparing all edges between the mock-up graph $\process^{\textit{tar}}$ and the implementation graph $\process$, we identify and output the inconsistent edges:
\begin{equation}\label{eq:process-change}
    \begin{aligned}
    \exists & t, (\screen_t^{\textit{tar}}, \actionchainset_t^{\textit{tar}}, \screen_{t+1}^{\textit{tar}}) \in \process^{\textit{tar}}, 
    (\screen_t, \actionchainset_t, \screen_{t+1}) \in \process  \textit{ such that } \textit{sim}(\screen_{t+1}, \screen_{t+1}^{\textit{tar}}) < \epsilon_{screen} 
    \end{aligned}
\end{equation}


\section{Approach}


\begin{figure*}[t]
    \centering
    \includegraphics[width=\textwidth]{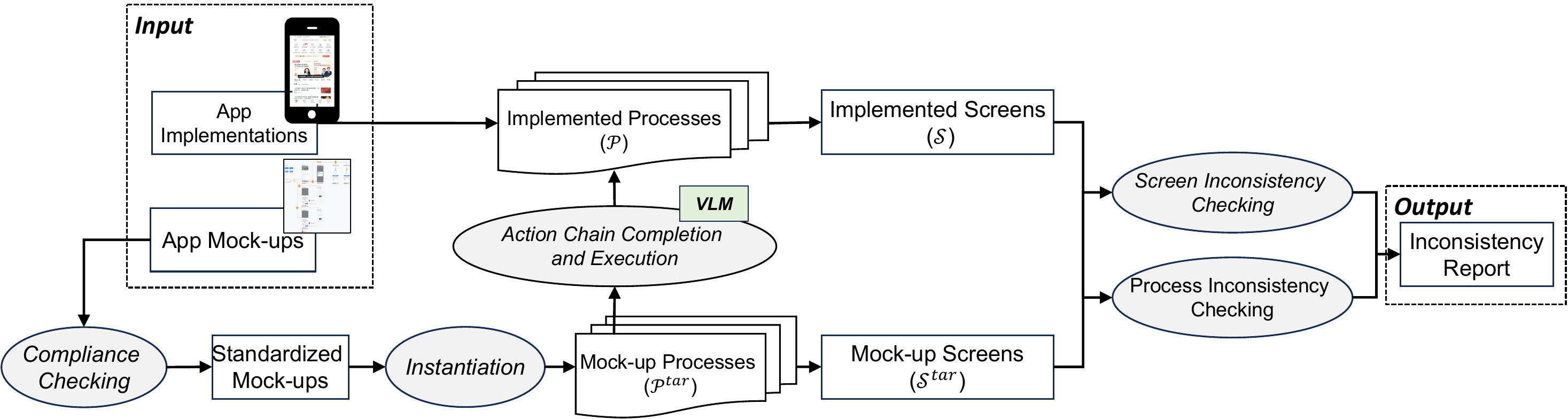}
    \caption{Overview of the \tool framework.
    It consists of three main modules: Mock-up compliance checking (Section \ref{app:meta-model}), Screen inconsistency checking (Section \ref{app:screen}), and Process inconsistency checking (Section \ref{app:process}).
    }
    \label{fig:overview}
\end{figure*}

\noindent\textbf{Overview.}
\autoref{fig:overview} presents the workflow of \tool.
The inputs consist of the implemented app running on a mobile device and a set of mock-up documents.
Each mock-up describes a scenario, such as ``buying exchange-traded funds (ETFs)''.

First, we perform a compliance check on these raw mock-ups to parse them, ensuring each process is reformulated into a standard meta-model format (Section \ref{app:meta-model}).
Based on the action chains specified in each process, we then attempt to execute the actions on the implementation (Section \ref{app:process}).
During this step, we utilize the VLM to auto-complete any implicit actions, ensuring the successful execution of the action chains.
With both the mock-up processes and the executed processes on the implementation, we can perform process inconsistency checks (Section \ref{app:process}).
Additionally, we compare all individual screen inconsistencies (Section \ref{app:screen}).
The final output is a report listing all the design violations in the implementation.

\subsection{Mock-up Meta-model and Compliance Checking}\label{app:meta-model}

\begin{figure}[h]
    \centering
    \includegraphics[width=0.5\textwidth]{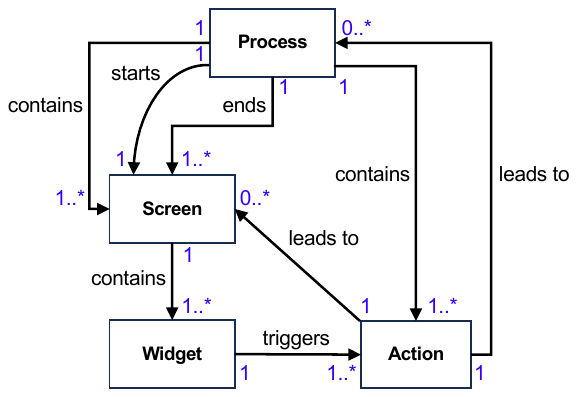}
    \caption{Meta-model for a design mock-up.}
    \label{fig:meta-model}
\end{figure}


As shown in \autoref{fig:meta-model}, our meta-model comprises four key elements: process, screen, widget, and action.
Each process features a unique starting screen and at least one ending screen.
A process can be represented as a connected directed graph consisting of multiple screens and the actions connecting them.
The graph can also be cyclic, where the starting screen is the same as the ending screen, such as when a user navigates to another screen and then returns to the initial one.
Actions are executed on the widget elements within each screen.
In this work, we consider the following action space:
click, long press, send keys, scroll, swipe, drag and drop, and go back.
When an action is performed, several outcomes are possible:
(i) it may result in no screen transition,
(ii) it may lead to another screen within the same process, or
(iii) it may lead to another process.
Note that, if a design mock-up does not conform to our meta-model,
\tool, as a tool, raises a compilation error for the designers to refine their mock-ups.

\subsection{Screen Inconsistency Detection}\label{app:screen}

In this work, we first learn a computer vision model to recognize our predefined widget types on the screens in both the design mock-ups and the application implementation.
For the detected widgets, we formulate an optimizable widget alignment problem and solve it with a dynamic programming solution.

\subsubsection{Widget Detection}\label{app:widget-detection}
Inspired by the previous work \cite{chen2020object, xie2020uied, lin2021phishpedia, bernal2022translating}, we adopt a vision-based approach to recognize widgets on both the mock-up screenshot and the implementation screenshot.
By training a state-of-the-art object detection model \cite{yolov8}, we can accurately identify widgets and their types as defined in Section \ref{sec:ps}.
This method unifies the widget extraction process for both the mock-up and the implementation, making it more generalizable to a wide range of applications.
For widget types that contain text (i.e., TextButton, CombinedButton, and TextView), we utilize an Optical Character Recognition (OCR) model to additionally extract the text content for later consistency checking.
As a result, a screen is converted to a set of widgets specified with their location, shape, and type as specified in Section~\ref{sec:ps}.


\begin{algorithm}[t]
\caption{Widget Matching Algorithm $f(.)$}\label{alg:widget-matching}
\SetKwInOut{Input}{Input}
\SetKwInOut{Output}{Output}
\setcounter{AlgoLine}{0}
\small

\Input{
Mock-up screen widget set $\widgetset_1 = \{\widget_i | \widget_i = (x_i, y_i, w_i, h_i, t_i)\}$, and the implementation screen widget set $\widgetset_2 = \{\widget_j | \widget_j = (x_j, y_j, w_j, h_j, t_j)\}$. 
}
\Output{
Matched pairs between two widget sets $\widgetset_1^{m}$, $\widgetset_2^{m}$
}

\BlankLine
$\widgetset_1^{m} \gets \emptyset$, $\widgetset_2^{m} \gets \emptyset$

\BlankLine
\tcp{Step 1: Sort widgets by partial ordering}
Sort $\widgetset_1$ such that $(y_i, x_i) \leq (y_{i'}, x_{i'}), \forall i<i'$ \\
Sort $\widgetset_2$ such that $(y_j, x_j) \leq (y_{j'}, x_{j'}), \forall j<j'$

\tcp{Step 2: Compute the similarity matrix}
Initialize similarity matrix $A \gets \mathbf{0}_{|\widgetset_1| \times |\widgetset_2|}$;
\BlankLine
\For{$\widget_i \in \widgetset_1$, $\widget_j \in \widgetset_2$}{
    \begin{align*}
       & sim_{pos} = min(\frac{1}{\alpha(|x_i - x_j| + |y_i - y_j|) + |w_i - w_j| + |h_i - h_j|}, 1) \\
       &  sim_{area} = \frac{\min \left( w_ih_i, w_jh_j \right)}{\max \left( w_ih_i, w_jh_j \right)} \\
       &  sim_{shape} = \frac{\min \left( w_i/h_i, w_j/h_j \right)}{\max \left( w_i/h_i, w_j/h_j \right)} \\
       &  sim_{type} = \mathbf{1}(t_i = t_j) + \delta \mathbf{1}(t_i \neq t_j), \textit{ where $0 < \delta < 1$} \\
       & A_{i,j} \gets sim_{pos} * sim_{area} * sim_{shape} * sim_{type};
    \end{align*}
}

\tcp{Step 3: LCS-based matching}
Initialize the matching matrix $M \gets \mathbf{0}_{(|\widgetset_1|+1) \times (|\widgetset_2|+1)}$;
\BlankLine
\For{$i=2,...|\widgetset_1|$, $j=2,...|\widgetset_2|$}{
    \begin{align*}
        M_{ij} \gets \max \{ M_{i,j-1} \;, M_{i-1,j} \;, M_{i-1,j-1} + A_{i-1,j-1}\} ;
    \end{align*}
}

\tcp{Backtrace the matched pairs}
$i \gets |\widgetset_1|, j \gets |\widgetset_2|$;
\BlankLine
\While{$i>1$ \textbf{and} $j>1$}{
    \If{$M_{ij} = M_{i-1, j-1} + A_{i-1,j-1}$}{
        $\widgetset_1^{m} \gets \widgetset_1^{m} \cup \widget_{i-1}$; \\
        $\widgetset_2^{m} \gets \widgetset_2^{m} \cup \widget_{j-1}$; 
    }
    \ElseIf{$M_{ij} = M_{i-1,j}$} {
            $i \gets i - 1$;
    }
    \Else{
            $j \gets j - 1$;
    }
}

\Return $\widgetset_1^{m}$, $\widgetset_2^{m}$;

\end{algorithm}

\subsubsection{Widget Alignment}\label{app:widget-matching}
We design a widget-matching algorithm as a dynamic programming problem shown in Algorithm \ref{alg:widget-matching}.
The inputs to the algorithm are two sets of widgets from the mock-up and the implementation and the output is the set of matched pairs between these two sets.
The algorithm is comprised of three critical steps:
\begin{itemize}[leftmargin=*]
   \item \textbf{Step 1: Widget \textit{Partial Order} (Lines 2-3 in Algorithm \ref{alg:widget-matching}).}
    We define \textit{partial order} for widgets appearing on the screen. Specifically, widgets are primarily sorted by their y-coordinate in ascending order (from top to bottom). For widgets appearing in the same row, they are secondarily sorted by their x-coordinate in ascending order (from left to right). 
    This partial ordering is derived based on our empirical observation that
    most widgets are positioned in a horizontal way\footnote{\tool adopt this partial order by default, but practitioners can customize their own partial order based on the needs.}.

    \item \textbf{Step 2: Widget Similarity Computation (Line 4-7 in Algorithm \ref{alg:widget-matching}).}
   We compute the pairwise similarities between the two sets of widgets.
   Our similarity metric includes four components:
   (i) position similarity is calculated using the L1-norm distance between the $(x, y, w, h)$ coordinates of the widgets, as originally adopted in \cite{moran2018automated},
   (ii) area similarity measures the differences in sizes ($w \times h$),
   (iii) shape similarity assesses the differences in aspect ratios ($\frac{w}{h}$),
    and (iv) type similarity assigns a score of 1 if the widgets share the same class.
    If the widgets are of different classes (e.g., input box vs. text button), the score is down-weighted by a factor of $\delta$.
    All four metrics are multiplied together to obtain the final similarity score.

    \item \textbf{Step 3: LCS-based Matching (Line 8-20 in Algorithm \ref{alg:widget-matching}).}
    Observing that the widgets on a screen can be treated as a sequence, 
    we convert the widget matching problem as a Longest Common Subsequence (LCS) problem \cite{lcs}.
    Our goal is to find the \textit{global} optimal sub-sequences of widgets from each set that achieve the highest cumulative similarity when matched.
    This is accomplished via dynamic programming \cite{dp}, resulting in two matched widget sets, $\widgetset_1^{m}$ and $\widgetset_2^{m}$.
    By framing the widget matching as an optimization problem, we can identify the best correspondence even in the presence of extra or missing widgets.

\end{itemize}

\subsubsection{Inconsistency Report Generation}\label{app:screen-inconsistency-report}
Based on the matching results, we can identify inconsistencies in screen implementation as defined in Section \ref{sec:ps}.
We consider three types of violations: extra widget, missing widget, and semantic change.
The first two types, i.e., extra and missing widgets, 
are identified by comparing the matched sets $\widgetset_1^{m}$ and $\widgetset_2^{m}$ with the initial sets $\widgetset_1$ and $\widgetset_2$.
The last type, i.e., semantic change, is computed on the matched widget pairs.
If the matched pair consists of different widget types, it is considered as a change.
For pairs sharing the same widget type, 
we follow the practice of GVT \cite{moran2018automated} to decide whether a widget is changed:
\begin{enumerate}[leftmargin=*]
    \item For text-based widgets (TextView, TextButton, InputBox, and CombinedButton), their text values sequence similarity ratio needs to be above a threshold $\epsilon_{ed}$.
    \item For icon-based widgets (IconButton, ImageView, InputBox, and CombinedButton),
    the binary color space difference must be below a threshold of $\epsilon_{binary}$,
    and the top-k most frequently occurring colors' RGB difference must also be within a threshold $\epsilon_{color}$.
\end{enumerate}

Last, we relax the comparison for the widget type ``Chart'', 
i.e., we consider two charts unchanged as long as they are aligned.
It is because the chart in the design mock-ups can be just an example,
its content can always change in the implementation.

\subsection{Process Inconsistency Detection}\label{app:process}
\begin{figure}[t]
    \centering
    \tiny
    \includegraphics[width=0.5\textwidth]{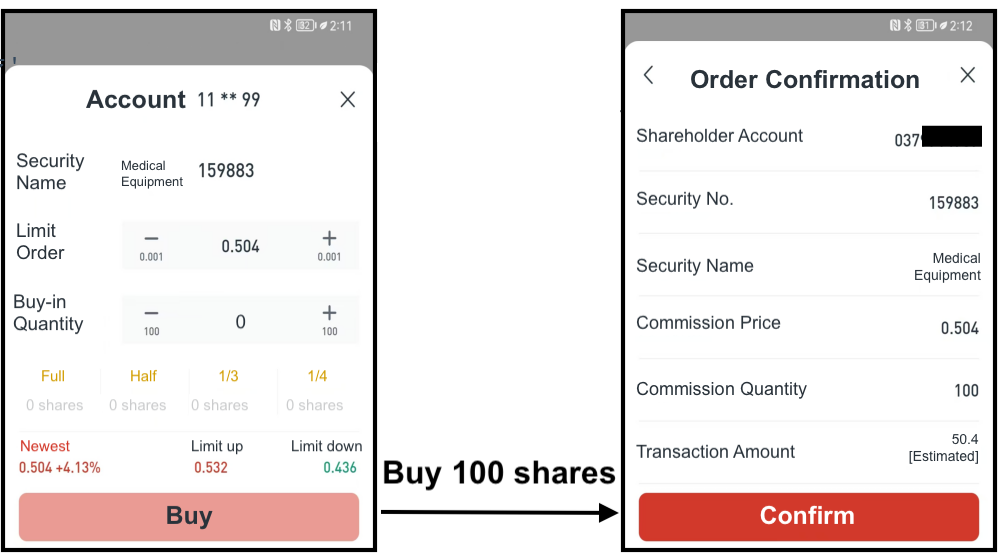}
    \caption{Security buying process, taken from a design mock-up of an industrial trading application.
    The transition is described in an abstract way as ``Buy 100 shares''.}
    \label{fig:process-eg}
\end{figure}

For each transition (in terms of a source screen $\textbf{s}_{src}$, a target screen $\textbf{s}_{tar}$, and their transition description $desc$)
in the design mock-up (as shown in \autoref{fig:process-eg})
and the source screen in the application $\textbf{s}^*_{src}$, we 
(1) translate $\textbf{s}_{src}$ and $desc$ to a sequence of actions on $\textbf{s}^*_{src}$ and
(2) compare the resulted screen in the application to see whether $\textbf{s}^*_{tar}$ is equivalent to the expected screen $\textbf{s}_{tar}$.

\begin{figure}[t]
    \centering
    \includegraphics[width=\textwidth]{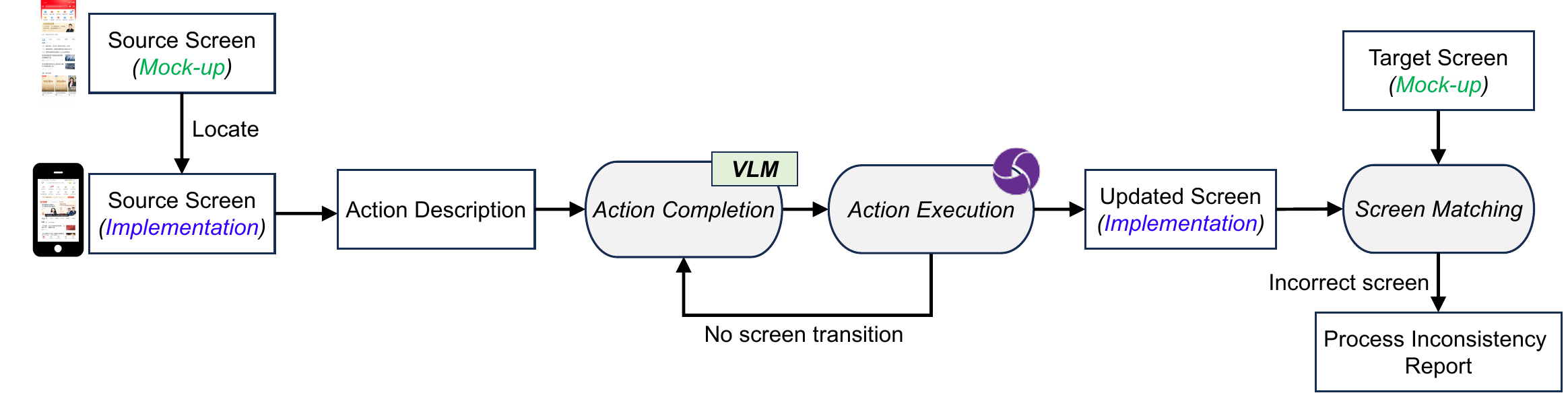}
    \caption{Process Execution Workflow.}
    \label{fig:process-inconsistency}
\end{figure}

Observing the ambiguous textual description and multi-modal design (i.e., text and image) as shown in \autoref{fig:process-eg},
we adopt the vision-language model to derive the concrete action sequence as shown in \autoref{fig:process-inconsistency},
which can be executed by the \textit{uiautomator} tool \cite{uiautomator2}.
If a screen transition occurs, we verify whether the updated screen matches the target screen in the mock-up.
A process inconsistency is reported when they do not match.
If the updated screen is correct and the process has follow-up actions, the updated screen becomes the new starting screen.

\subsubsection{Visual Prompt Design}\label{app:visual-prompt}

\begin{table*}[t]
    \centering
    \scriptsize
    \caption{
    Visual prompt for process inconsistency checking with action completion.
    The components in blue are mutable and vary according to the current screen content.}
    \label{tab:vlm-prompt}
    \begin{tabular}{p{2.5cm}|p{10cm}}
        \toprule
        \multicolumn{2}{l}{\textbf{System Prompt (Not overwritable)}} \\
        \midrule
        \coloredbox{SoftRed}{\textbf{Task Objective}} & 
        Given the current GUI screen, you need to return a sequence of actions to transit to the next screen.
        \\
        \midrule
        \coloredbox{SoftPurple}{\textbf{I/O Description}} & 
        The input is the \textbf{current GUI screenshot} with annotations for the widget bounding boxes (actionable widgets).
        
        The output should be a \textbf{list of actions}. 
        Actions can be one of the following: 
        \begin{enumerate}
            \item \textbf{click}(widget\_id): This includes activating an input box, toggling a switch, checking a checkbox, etc.
            \item \textbf{long\_press}(widget\_id)
            \item \textbf{send\_keys}(value): Once a widget is selected, one can set the value of it.
            \item \textbf{scroll}(widget\_id, direction, distance): Scroll to the left, right, up, or bottom by some pixels of distance. If widget\_id is not specified, the default operation is to scroll the entire screen.
            \item \textbf{swipe}(widget\_id, direction)
            \item \textbf{drag\_and\_drop}(widget\_id1, widget\_id2): Drag the widget\_1 to the center of widget\_2.
            \item \textbf{go\_back}(): This would return to the previous screen.
        \end{enumerate}
        \\
        \midrule
        
        \coloredbox{SoftGreen}{\textbf{Few-shot Example}} & 
        For example, if we have the current GUI screen with two widgets: 
        
        widget\_1 is a button with text ``confirm'', widget\_2 is an input box with placeholder text as ``please input the password''. 
        
        After I perform \textbf{click(widget\_1)} action on the current GUI screen,
        the screen does not change, this indicates that widget\_2 is unfilled.
        
        Therefore, the revised action chain should be 
        \textbf{click(widget\_2)},
        \textbf{send\_keys(``my\_password'')}, 
        \textbf{click(widget\_1)}. 
        \\

        \midrule
        \multicolumn{2}{l}{\textbf{User Prompt (Modifiable)}} \\
        \midrule
        \multicolumn{1}{p{2.5cm}}{  
            \vbox{\vspace{0pt}\includegraphics[width=\linewidth]{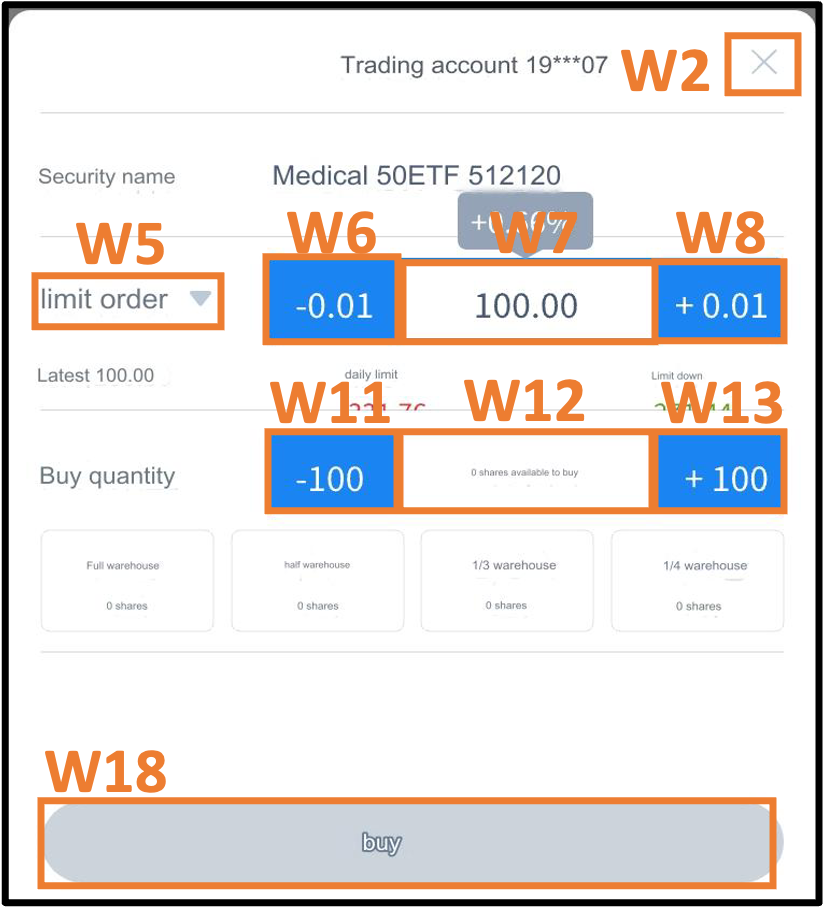}\vspace{0pt}}
        }  &
        \multicolumn{1}{p{10cm}}{
            \vbox{
                \vspace{0pt}

                Given the current screen and description, 
                please provide the next immediate correct action(s).
                
                \coloredbox{SoftBlue}{\textbf{Action Input}} 
                
                \{Action description in Natural language. E.g. ``Buy 100 shares and proceed''\}

                \coloredbox{SoftBlue}{\textbf{GUI Screenshot Input}} 
                
                \{Current GUI screenshot as shown in the left.\}

                \coloredbox{SoftBlue}{\textbf{(Optional) Feedback}} 
                
                Your previous answer is incorrect, 
                are you sure you are referring to the correct widget, 
                or does the action exist in the action space? 
                \vspace{0pt}
            } 
        } \\
        
        \bottomrule
    \end{tabular}
\end{table*}

We use VLM to translate ``ambiguous design'' to ``standard action''.
For example, given an ambiguous transition description like ``trigger the submit button'', 
VLM can look into the GUI screen with widget IDs, 
and retrieve the widget ID corresponding to the submit button (e.g., widget 1).
Lastly, it returns an executable action in the format ``click(widget\_1)''.
This process incurs two major challenges:

\begin{itemize}[leftmargin=*]
  \item \textit{Noisy information on the screen}:
    A screen can contain a myriad of elements, 
    including both interactable and non-interactable widgets, as well as app-irrelevant icons.
    Irrelevant icons can be distracting in the action completion task and should therefore be ignored.
  \item \textit{Parsability and flexibility of VLM response}:
    VLMs typically generate responses in a very flexible manner,
    which is hard to parse into programmable actions on the mobile application.
\end{itemize}

To this end, we propose a visual prompt for VLM to focus on the crucial widgets and restrict its response into a parsable format.
Table \ref{tab:vlm-prompt} details our visual prompt,
consisting of a system prompt shared across all samples and a user prompt that is customized based on the current GUI screen.
On the one hand, we use the widget detection model (see Section~\ref{app:widget-detection})
to highlight the interactable widgets only, attached with their widget indices.
By this means, VLM can have more explicit areas to focus on and generate responses with reference.
On the other hand, 
we define the action space of LLM, which covers the major types of widget interactions: click, long press, send keys, scroll, swipe, and drag.
Note that we mainly consider app-related interactions, not system-level interactions such as toggling the Wifi, Bluetooth, etc.
To further control the randomness, we take an in-context learning approach.
Specifically, we provide a few-shot example that showcases the expected response format.
To save the token usage, the few-shot example is described in plain text.

In the user prompt, the query consists of both text and image modalities.
The action description in natural language is fed into the LLM.
And the query screenshot is annotated with interactable widgets and their IDs, this can instruct the VLM to pay special attention to the highlighted widgets only.
The VLM query process can be iterative until a screen transition happens or an interaction limit is reached.
If the screen has no change, we will provide feedback to VLM and ask it to reflect on its previous answer.

\subsubsection{Screen Matching}\label{app:screen-matching}
Once the screen has been navigated, we need to verify whether the transited screen matches the target screen in the mock-up.
Our design of the screen matching metric is based on the widget matching results in Algorithm \ref{alg:widget-matching}.
Given the updated screen and target screen, we compute the sum of similarities for all matched pairs, normalized by the total number of widgets in the target screen (\autoref{eq:screen-similarity}).
By applying a threshold to $\textit{sim}(\screen, \screen^{\textit{tar}})$, we can determine whether the updated screen is sufficiently close to the target screen.

\begin{equation}\label{eq:screen-similarity}
\resizebox{0.7\textwidth}{!}{
    $\textit{sim}(\screen, \screen^{\textit{tar}}) = \frac{\underset{\widget_i \in \widgetset, \widget_j \in \widgetset^{tar}}{\sum}{ A_{i,j} \cdot \mathbf{1}(\widget_i \in \widgetset, \widget_j \in \widgetset^{tar}\textit{ and they matched})}}{|\widgetset^{tar}|}$
    }
\end{equation}

\subsubsection{Inconsistency Report Generation}\label{app:process-inconsistency-report}
As defined in Section \ref{sec:ps}, the expected process in the mock-up is denoted as $\process^{\textit{tar}} = (\screenset^{\textit{tar}}, \actionchainset^{\textit{tar}})$, whereas the executed process in the implementation is $\process = (\screenset, \actionchainset^{\textit{tar}}_{\textit{complete}})$.
We report a process inconsistency if any of the $\screenset$ does not match $\screenset^{\textit{tar}}$, i.e. $\exists \screen_t \in \screenset, \screen^{\textit{tar}}_t \in \screenset^{\textit{tar}}, \textit{sim}(\screen_t, \screen^{\textit{tar}}_t) < \epsilon_{screen}$.
Here, $t$ denotes the timestep, with $\epsilon_{screen}$ being the similarity threshold.  
\section{Experiments}

We evaluate \tool with the following research questions in mind:
\begin{itemize}[leftmargin=*]
    \item RQ1 (Screen Consistency Experiment):
    How effectively does \tool detect \textit{screen inconsistencies} in the public mobile application dataset?
    \item RQ2 (Process Consistency Experiment):
    How effectively does \tool detect \textit{process inconsistencies} in the public mobile application dataset?
    \item RQ3 (Component-wise Evaluation): 
    How accurate are the critical components in \tool? Specifically,
    \begin{itemize}
        \item RQ3-1 What is the performance of the \textit{widget detection} model?
        \item RQ3-2 Can VLM successfully convert transition descriptions into executable actions?
    \end{itemize}
\end{itemize}

\noindent\textbf{Datasets.}
We collect 80 of the top free public applications from Google Play, 
from the categories of Business, Communication, Finance, and Social.
We recruit four experts, each possessing at least two years of software development experience.
For each app, the experts were asked to manually label two application scenarios as the mock-up processes, such as ``following a social account'' and ``setting notification preferences''.
They record all the screens and intermediate action chains associated with these processes.
These annotations serve as our simulated mock-ups.
We derive two design mock-ups from each application, 
and each design mock-up comprises on average eight screens.
Interested readers can refer to \cite{dataset-examples} for more examples of collected processes.

\noindent\textbf{Configuration.}
The hyperparameters $\alpha$ and $\delta$ in Algorithm \ref{alg:widget-matching} are selected through grid search. 
We use the screen consistency experimental performance (as introduced in Section \ref{exp:rq1}) as the metric to guide the selection process. 
The best $\alpha$ is chosen to 10. 
This scaling factor ensures that the range of $sim_{pos}$ is more evenly distributed across (0,1]. 
Without this scaling, $sim_{pos}$ tends to cluster around 1 when the widget distance is small, reducing its discriminative power. 
Additionally, $\delta$ is set to 0.5, meaning that when two widgets belong to different classes, 
their final similarity is reduced by half.

The following hyperparameters are directly adopted from GVT \cite{moran2018automated}.
Specifically, we consider a string similarity ratio below $\epsilon_{ed} = 0.95$ as a text violation.
When calculating color differences, we extract the Top 3 most frequently occurring colors from each widget and compute their RGB differences.
A color difference exceeding $\epsilon_{color} = 0.05$ is considered a color violation.
Additionally, the proportion of pixels that differ in binary color space is restricted to no more than $\epsilon_{binary} = 20\%$.

Regarding process inconsistencies, we choose the optimal screen matching threshold ($\epsilon_{screen}$ in Section \ref{app:process-inconsistency-report}) using the 100 mutated design mock-ups introduced in the process consistency experiment (Section \ref{exp:rq2}).
The threshold is selected to be 0.73, which achieves the highest F1-score.

\subsection{RQ1: Screen Consistency Experiment}\label{exp:rq1}
\subsubsection{Setup}
We mutate the screen of the simulated mock-ups to inject screen inconsistencies.
Specifically, we choose the following GUI mutation types, 
which are common in the real application, 
covering 92\% of the mutation cases \cite{moran2018automated}.
It is a reasonable assumption that the majority of the widgets are correctly implemented.
Therefore, in each mutation case, we select 5\% of the widgets on the screen to be modified, which averages to 1-2 widgets per screen. This setup is the same as \cite{moran2018automated}.

\begin{itemize}[leftmargin=*]
    \item \textbf{Missing widgets:}
     For each screen, we randomly select 5\% of the widgets to delete. To simulate a realistic rendering effect, we remove the entire row containing the selected widget and shift the remaining widgets upward.
    \item \textbf{Extra widgets:}
    For each screen, we randomly insert approximately 5\% additional widgets. We add complete rows for these widgets and shift the existing widgets downward accordingly.
    \item \textbf{Semantic change -- Swapped widgets:}
    For each screen, we randomly select 5\% of the widgets and swap them with widgets of different types to introduce semantic changes.
    \item \textbf{Semantic change -- Text change:}
    For each screen, we randomly select 5\% of text-based widgets and alter their text content to create semantic inconsistencies.
    \item \textbf{Semantic change -- Color change:}
    For each screen, we randomly select 5\% of image-based widgets and alter their colors to introduce semantic discrepancies.
\end{itemize}

\subsubsection{Baseline}
We choose GVT \cite{moran2018automated} as our baseline as it is the state-of-the-art screen comparison solution for mobile applications.
We follow the configurations in \cite{moran2018automated} in this experiment.
To test the necessity of our model design, 
we also include a simple baseline of direct querying VLM for inconsistency checking.
Specifically, we feed the two screens, each annotated with widget IDs, to the VLM and prompt it to decide whether some of the widgets have been missed, inserted, or semantically edited. 

\subsubsection{Metrics}
Following the metrics used in \cite{moran2018automated}, 
we use precision, recall, Jaccard index, and classification precision as the evaluation metrics in this experiment.
Let $TP$, $FP$, $FN$, and $TP_c$ represent true positive inconsistency, false positive inconsistency, false negative inconsistency, and true positives with the correct type, respectively.
We calculate precision, recall, Jaccard index, and classification precision as follows:
\begin{equation}
    \begin{aligned}
        & pre = \frac{TP}{TP + FP}, & rec = \frac{TP}{TP + FN}\\
        & J\_Index = \frac{TP}{TP + FN + FP}, & 
        cp = \frac{TP_c}{TP} \\
    \end{aligned}
\end{equation}\label{eq:metrics}
Specifically, precision $pre$ is the number of reported true inconsistencies divided by the total number of reported inconsistencies.
Recall $rec$ is the number of reported true inconsistencies divided by the total number of inconsistencies.
The Jaccard Index $J\_Index$ punishes both false negatives (unreported real inconsistencies) and false positives (false alerts).
Classification precision $cp$ is the number of real reported inconsistencies correctly identified as the correct type divided by the total number of real reported inconsistencies.

\begin{table*}[t]
    \centering
    \caption{The results of screen consistency experiment}\label{tab:screen-inconsistency-results}
    \resizebox{\textwidth}{!}{%
    \begin{tabular}{llccccc}
    \toprule
    \textbf{Mutation Type} & \textbf{Solution} & \textbf{Precision} & \textbf{Recall} & \textbf{Classification Precision} & \textbf{Jaccard Index} & \textbf{Median Time (s)} \\
    \midrule
    \textbf{Extra} & \textit{GVT} & 0.793 & 0.899 & 1.000 & 0.690 & 0.001 \\
     & \textit{VLM} & 0.088 &	0.137 &	1.000 &	0.056 &	1.230 \\
      & \textit{GUIPilot} & \textbf{0.998} & \textbf{0.986} & 1.000 & \textbf{0.982} & 0.001 \\
    \midrule
    \textbf{Missing} & \textit{GVT} & 0.912 & 0.938 & 1.000 & 0.840 & 0.001 \\
     & \textit{VLM} & 0.123 &	0.154 &	1.000 &	0.073 & 1.430 \\
      & \textit{GUIPilot} & \textbf{0.997} & \textbf{0.984} & 1.000 & \textbf{0.978} & 0.001 \\
    \midrule
    \textbf{Swap} & \textit{GVT} & 0.283 & 0.430 & 1.000 & 0.200 & 0.001 \\
     & \textit{VLM} & 0.045 &	0.078 &	0.910 &	0.026 &	1.530 \\
      & \textit{GUIPilot} & \textbf{0.987} & \textbf{0.992} & 1.000 & \textbf{0.971} & 0.001 \\
    \midrule
    \textbf{Text Change} & \textit{GVT} & 0.998 & 0.999 & 0.981 & 0.960 & 0.001 \\
    & \textit{VLM} & 0.119 &	0.248 &	0.992 &	0.086 &	1.870 \\
     & \textit{GUIPilot} & 0.996 & 0.999 & 0.981 & 0.960 & 0.001 \\
    \midrule
    \textbf{Color Change} & \textit{GVT} & 1.000 & 0.999 & 1.000 & 0.990 & 0.001 \\
     & \textit{VLM} & 0.075 &	0.177 &	0.964 &	0.052 &	3.110 \\
      & \textit{GUIPilot} & 1.000 & 0.999 & 1.000 & 0.990 & 0.001 \\
    \bottomrule
    \end{tabular}
    }
\end{table*}

\begin{figure}[t]
    \centering
    \begin{subfigure}[t]{0.24\textwidth}
        \fbox{\includegraphics[width=\textwidth, keepaspectratio]{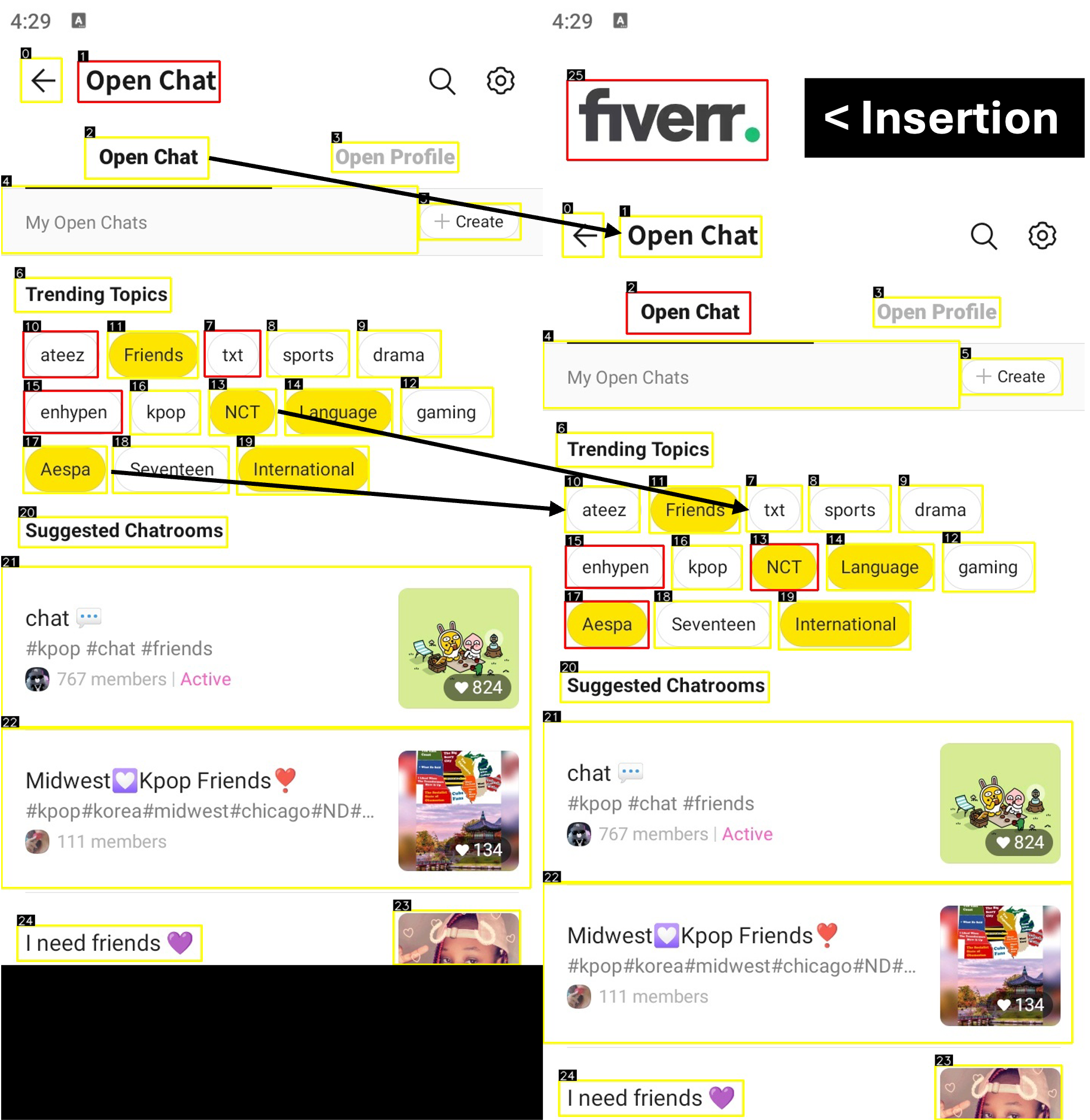}}
        \subcaption{Example 1: GVT}
    \end{subfigure}%
    \hfill
    \begin{subfigure}[t]{0.24\textwidth}
        \fbox{\includegraphics[width=\textwidth, keepaspectratio]{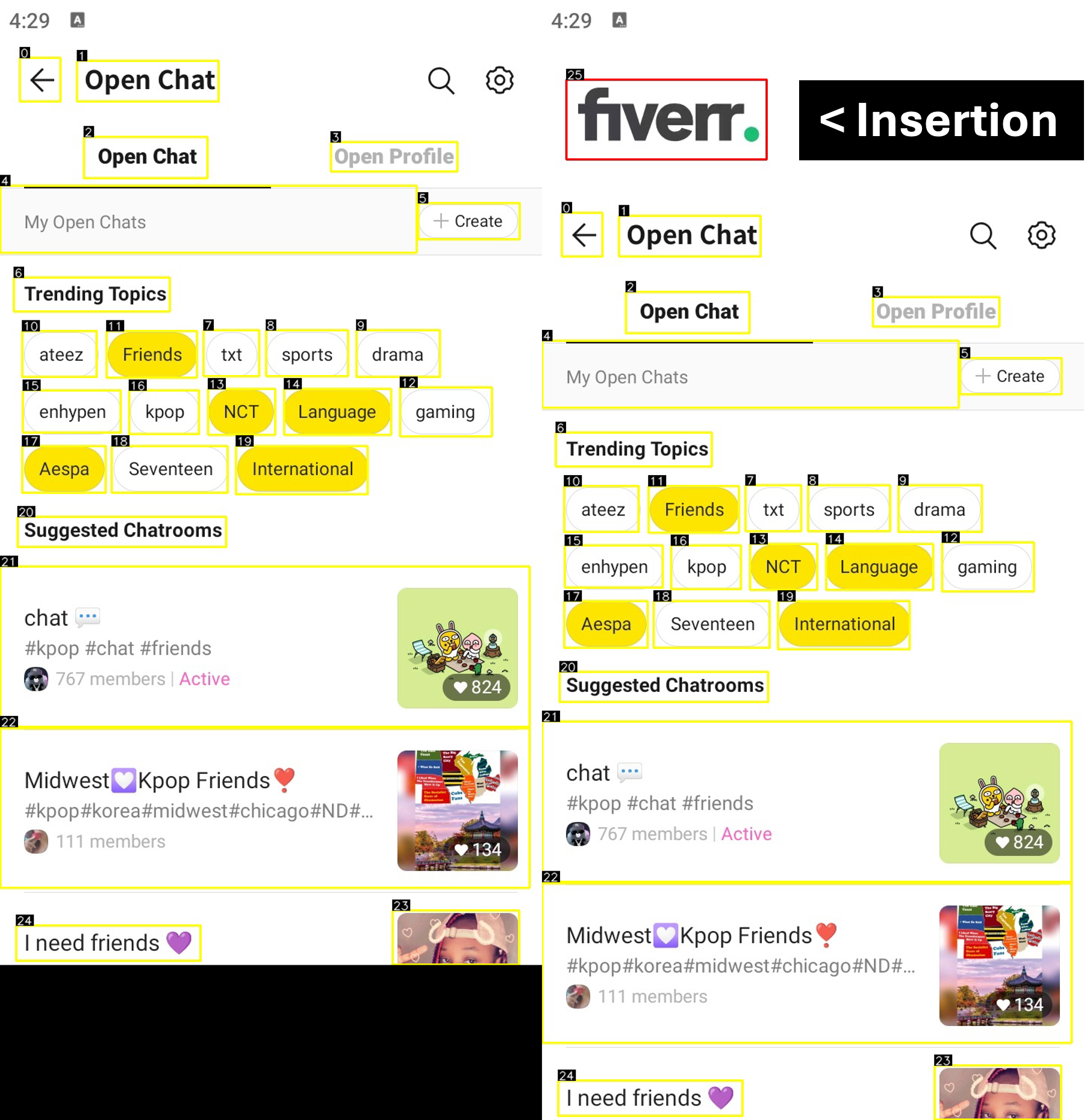}}
        \subcaption{Example 1: \tool}
    \end{subfigure}%
    \hfill
    \begin{subfigure}[t]{0.24\textwidth}
        \fbox{\includegraphics[width=\textwidth, keepaspectratio]{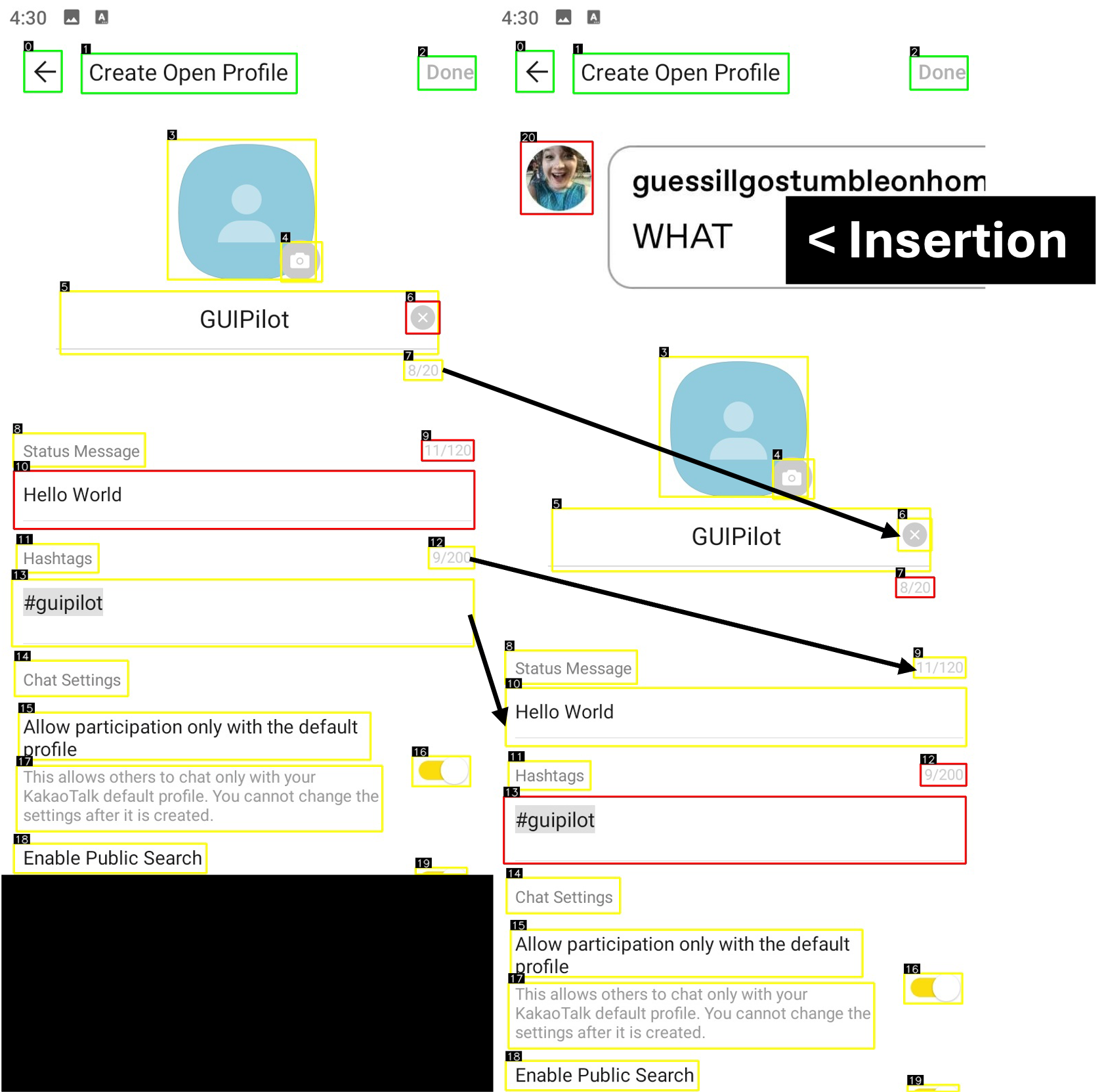}}
        \subcaption{Example 2: GVT}
    \end{subfigure}%
    \hfill
    \begin{subfigure}[t]{0.24\textwidth}
        \fbox{\includegraphics[width=\textwidth, keepaspectratio]{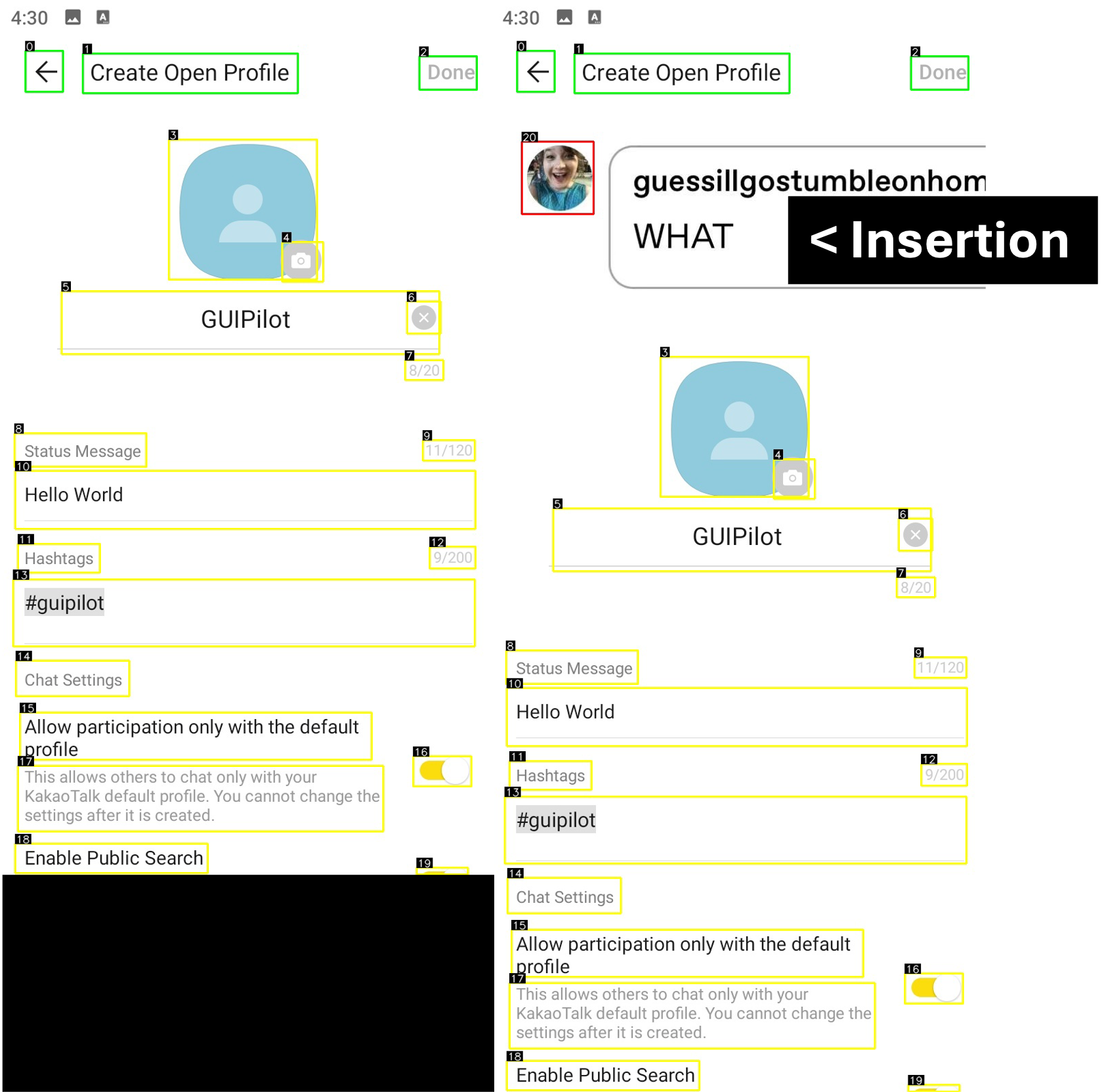}}
        \subcaption{Example 2: \tool}
    \end{subfigure}
    \vspace{1em}

    \begin{subfigure}[t]{0.24\textwidth}
        \fbox{\includegraphics[width=\textwidth, keepaspectratio]{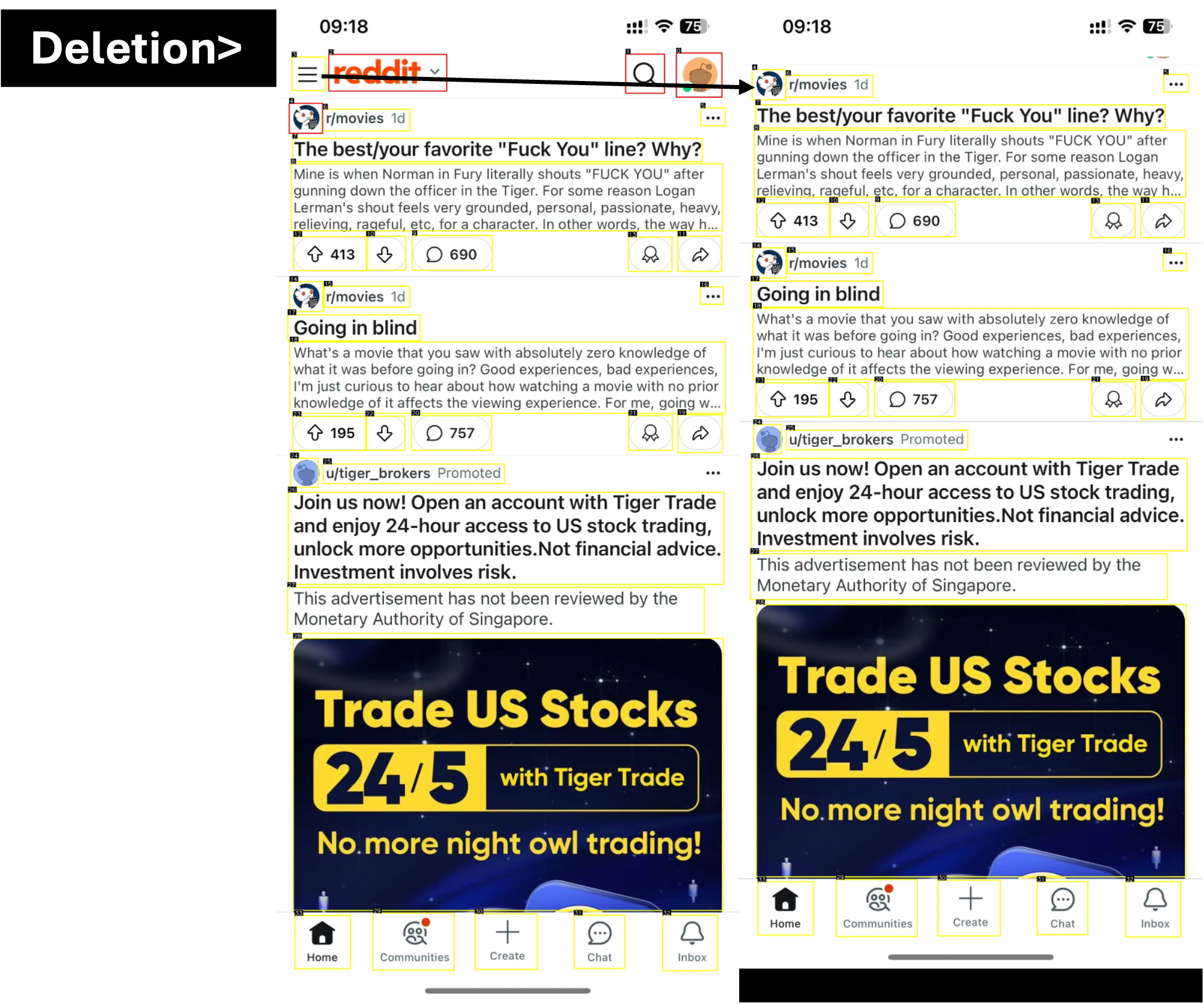}}
        \subcaption{Example 3: GVT}
    \end{subfigure}%
    \hfill
    \begin{subfigure}[t]{0.24\textwidth}
        \fbox{\includegraphics[width=\textwidth, keepaspectratio]{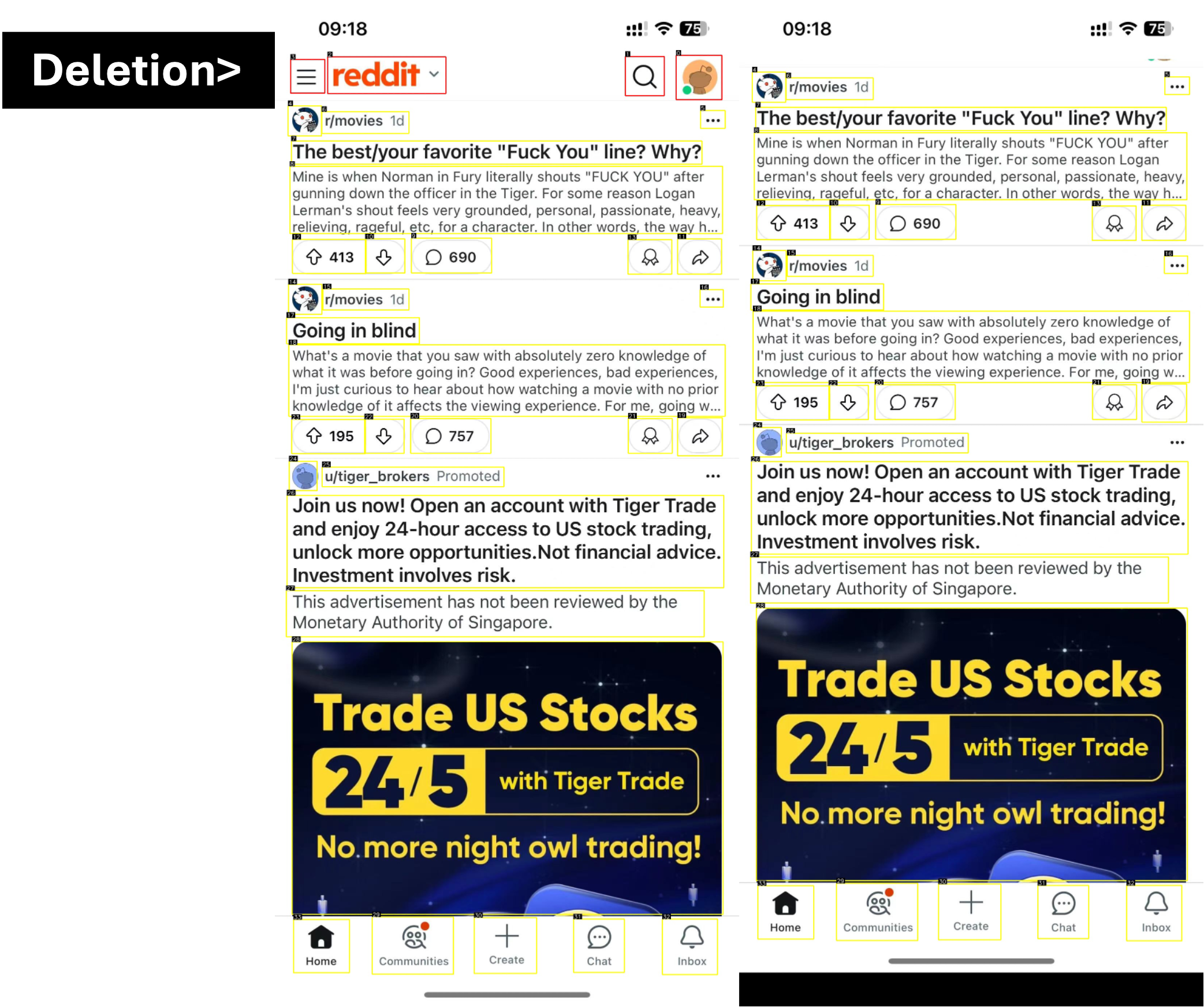}}
        \subcaption{Example 3: \tool}
    \end{subfigure}
    \begin{subfigure}[t]{0.24\textwidth}
        \fbox{\includegraphics[width=\textwidth, keepaspectratio]{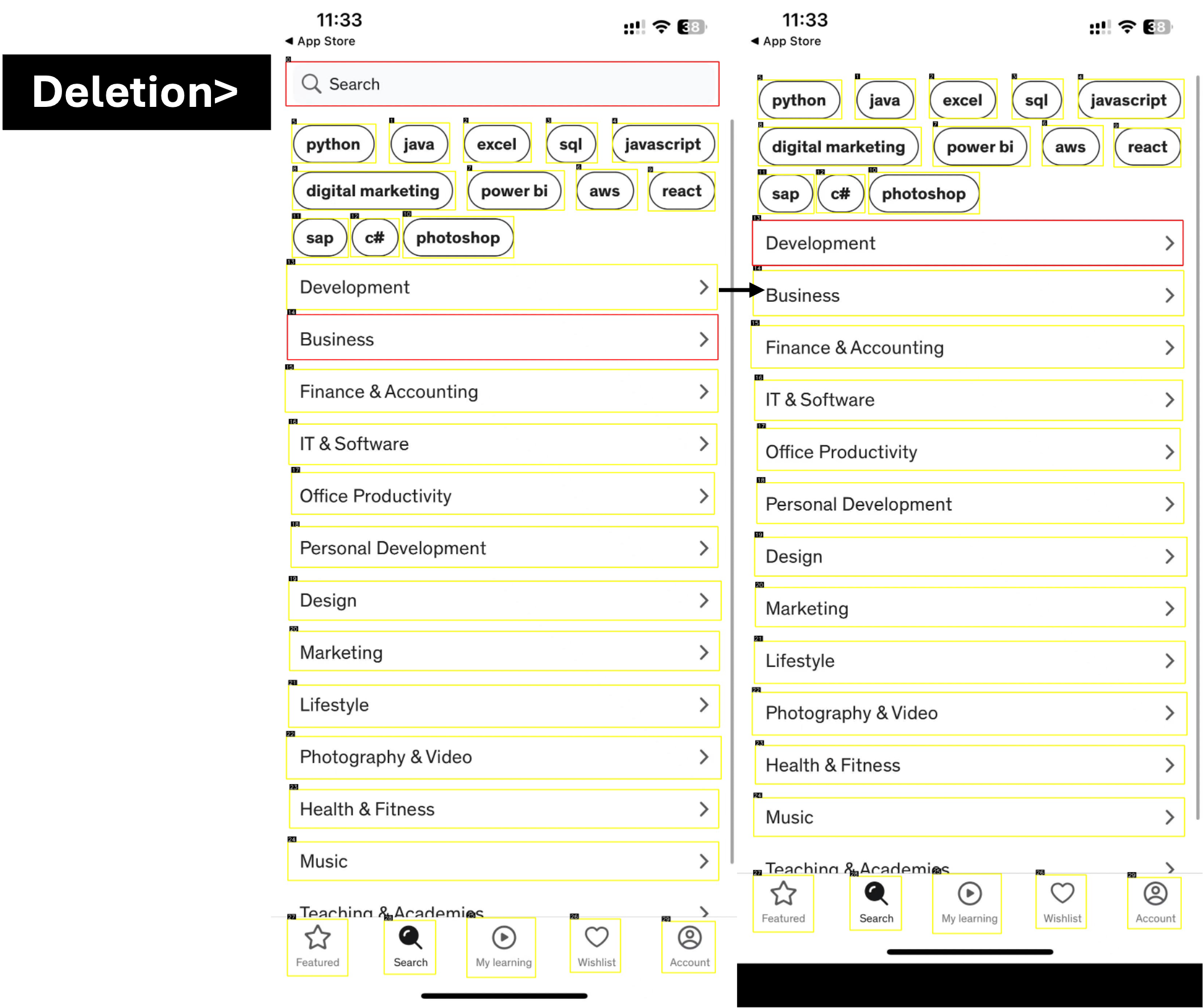}}
        \subcaption{Example 4: GVT}
    \end{subfigure}%
    \hfill
    \begin{subfigure}[t]{0.24\textwidth}
        \fbox{\includegraphics[width=\textwidth, keepaspectratio]{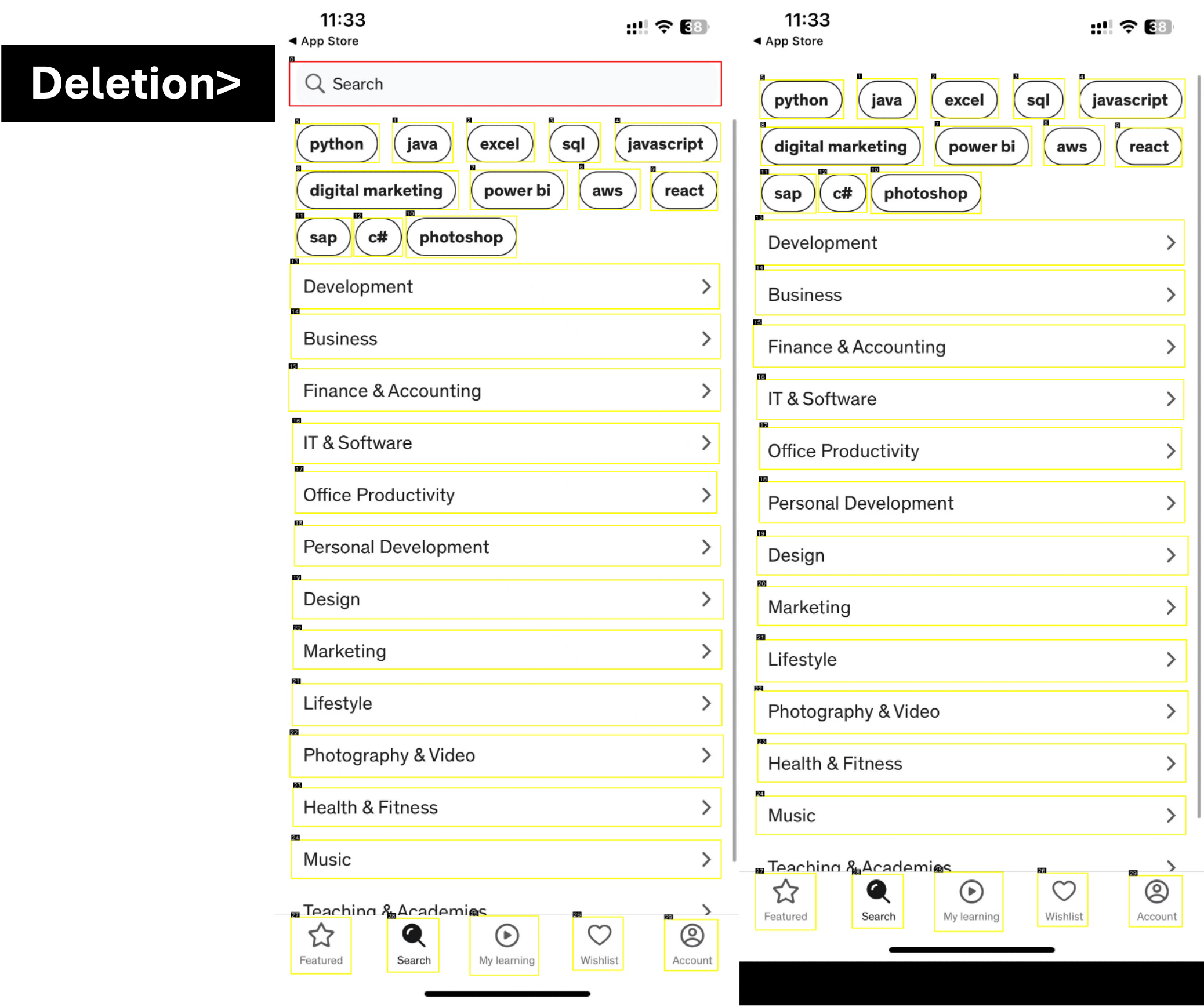}}
        \subcaption{Example 4: \tool}
    \end{subfigure}
    \vspace{1em}

    \begin{subfigure}[t]{0.24\textwidth}
        \fbox{\includegraphics[width=\textwidth, keepaspectratio]{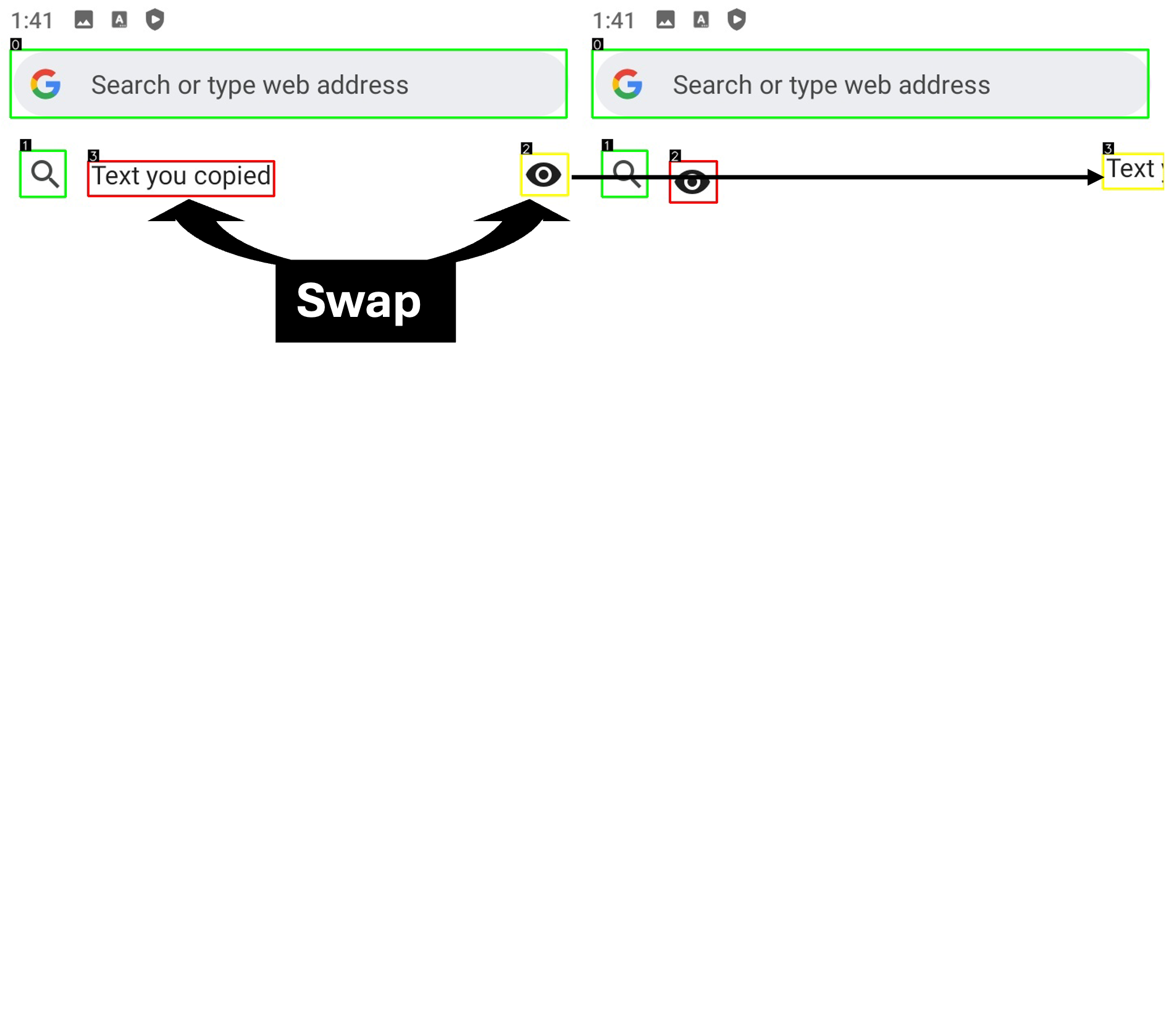}}
        \subcaption{Example 5: GVT}
    \end{subfigure}%
    \hfill
    \begin{subfigure}[t]{0.24\textwidth}
        \fbox{\includegraphics[width=\textwidth, keepaspectratio]{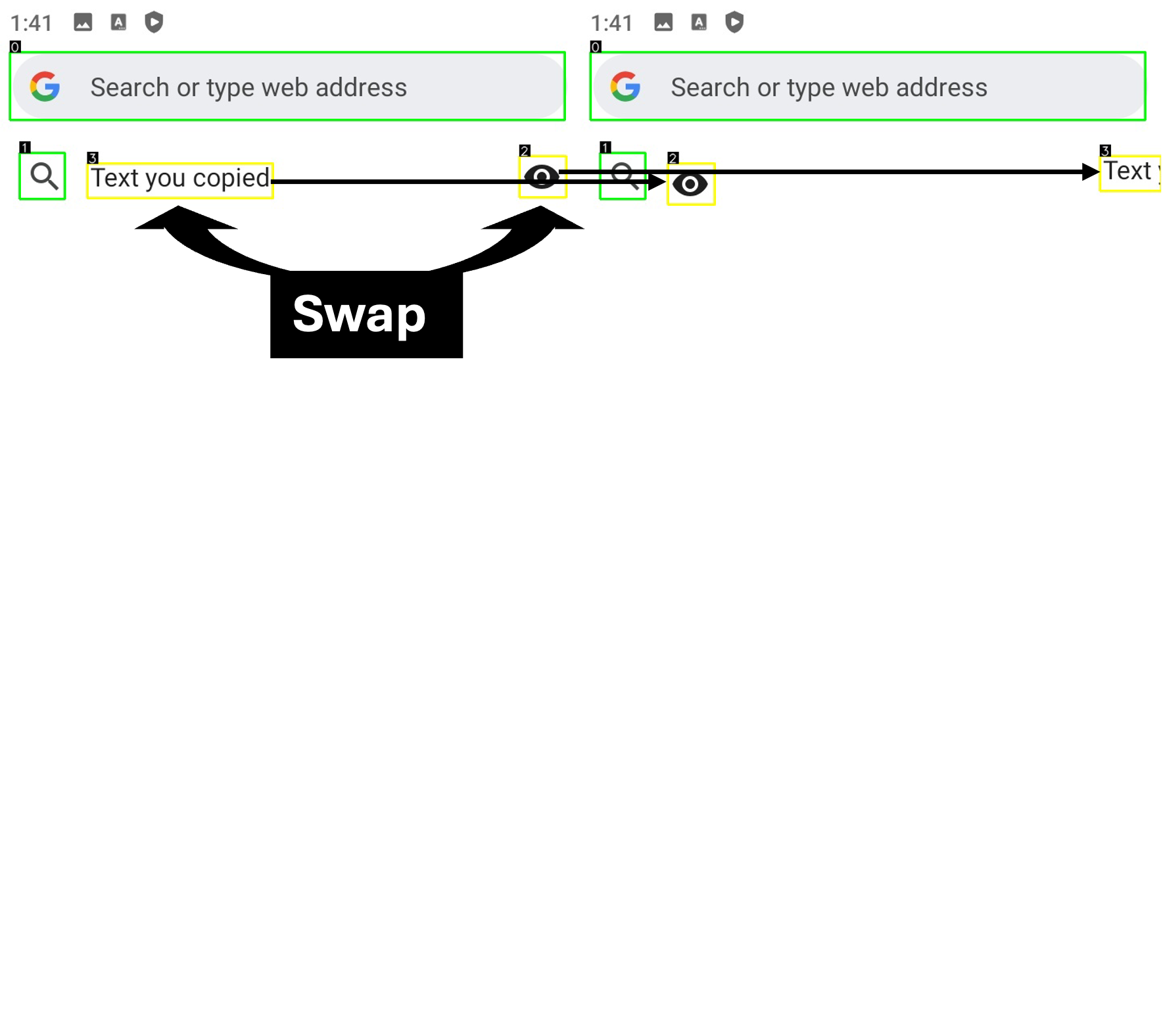}}
        \subcaption{Example 5: \tool}
    \end{subfigure}
    \hfill
    \begin{subfigure}[t]{0.24\textwidth}
        \fbox{\includegraphics[width=\textwidth, keepaspectratio]{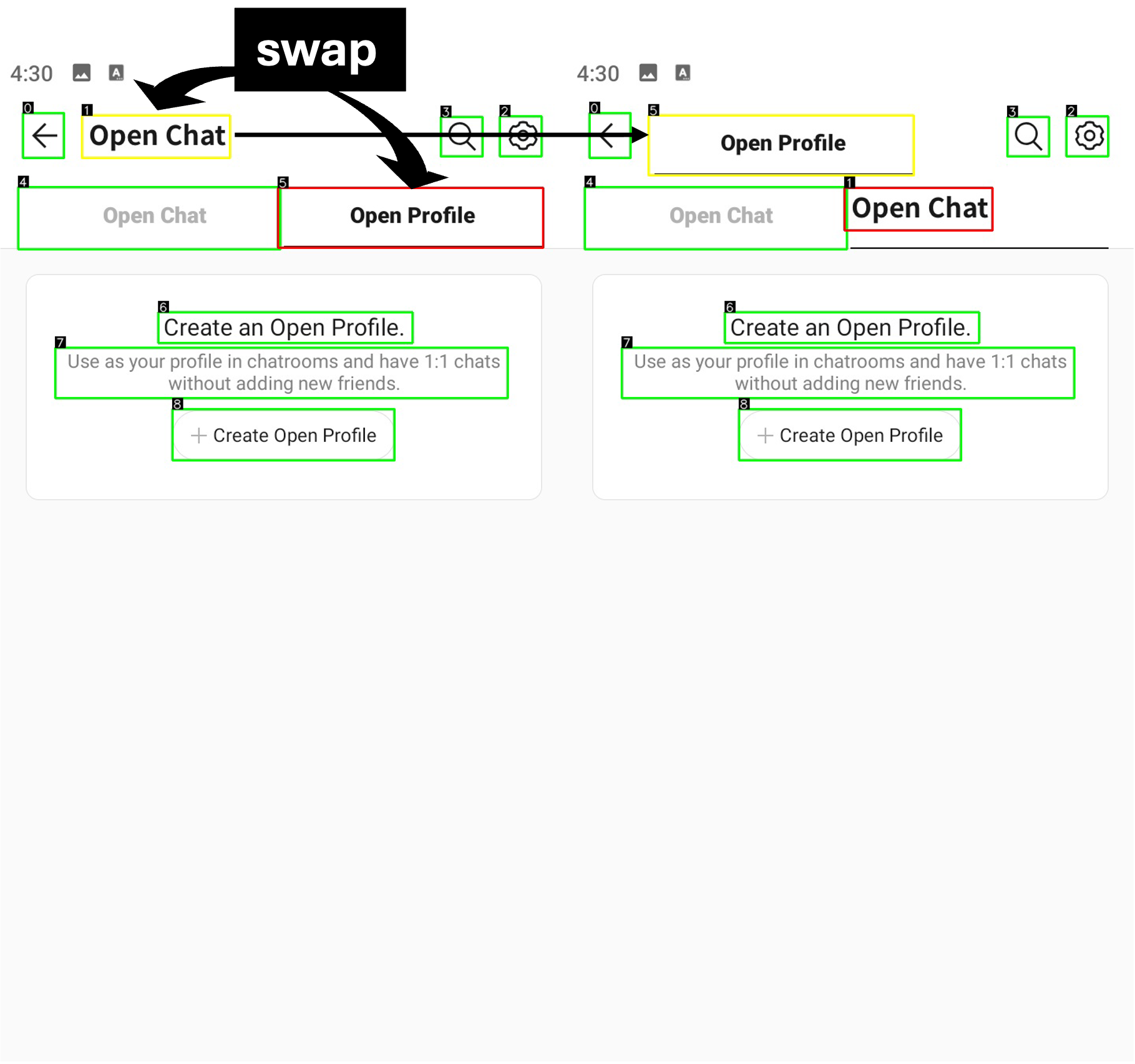}}
        \subcaption{Example 6: GVT}
    \end{subfigure}%
    \hfill
    \begin{subfigure}[t]{0.24\textwidth}
        \fbox{\includegraphics[width=\textwidth, keepaspectratio]{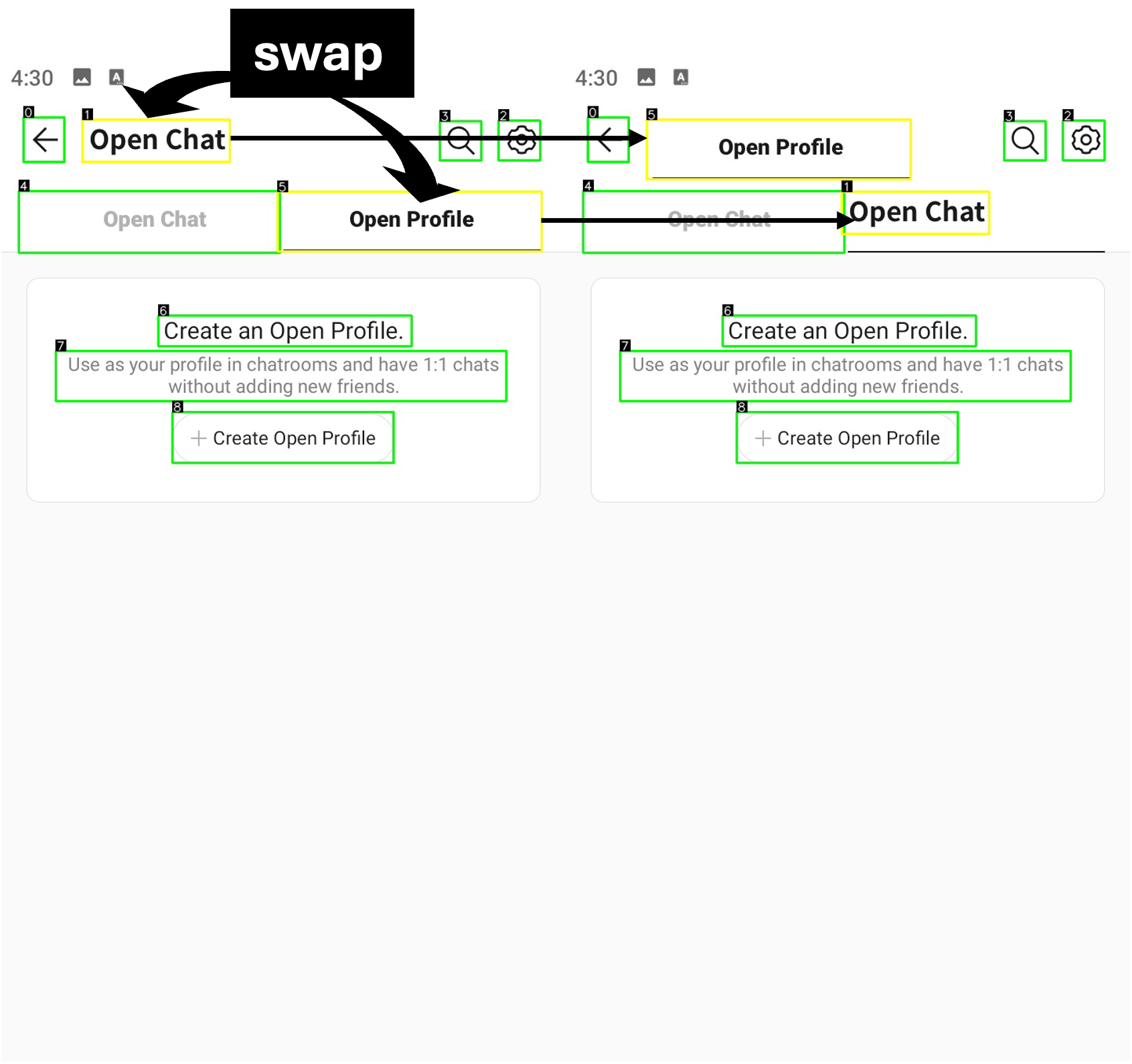}}
        \subcaption{Example 6: \tool}
    \end{subfigure}
    \caption{
    Comparison between GVT and \tool.
    In each figure, the original screen is displayed on the left, and the mutated screen (after insertion, deletion, or swapping) appears on the right.
    \coloredbox{SoftRed}{Red} boxes indicate extra or missing widgets.
    \coloredbox{SoftGreen}{Green} boxes denote widgets that are unaffected by the mutation.
    \coloredbox{SoftYellow}{Yellow} boxes are those who have shifted due to the mutation but can still identify a match.
    \textbf{Lines} highlight certain matches reported by the approach.}
    \label{fig:gui-win}
\end{figure}

\subsubsection{Results}
Table \ref{tab:screen-inconsistency-results} presents a comparative analysis between \tool and GVT \cite{moran2018automated}.
Overall, \tool shows better performance in detecting layout violations, such as extra widgets and widget swaps.
For in-place semantic changes such as color and text change, \tool's performance is on par with GVT.
Notably, \tool achieves these results without incurring additional runtime overhead.
Theoretically, our widget alignment implements the longest-common-subsequence matching, which incurs a time complexity of $\mathcal{O}(mn)$ where $m$ is the total number of widgets on the mock-up and $n$ is the total number of widgets on the implementation. 
Whereas GVT \cite{moran2018automated} identifies the nearest neighbor on the implementation screen for every widget on the mock-up, this has a time complexity of $\mathcal{O}(mnlogn)$.

Additionally, directly querying VLM results in poor classification performance, with unacceptably high runtime. 
We observe that VLM may hallucinate non-existent inconsistencies (false positives) or identify inconsistencies but assign incorrect widget IDs (false positives and false negatives). 
This limitation is likely due to the resolution constraints of the visual encoder and the limitations of the local attention mechanism \cite{liu2024ocrbench, song2025layertracer}.
We conduct a qualitative analysis of GUIPilot and GVT as follows, more examples can be found in our anonymous website \cite{qualitative-rq1}.

\paragraph{Why \tool is better than GVT?}
First, GVT is more vulnerable to location shifts.
GVT predominantly depends on the relative positions of widgets on the screen for matching, making it susceptible to minor shifts in the GUI layout.
In contrast, \tool takes into account not only the widget's location but also its shape and type, enabling it to correctly match corresponding widget pairs even when minor shifts occur.
As illustrated in Figure \ref{fig:gui-win}, examples 1-4 demonstrate that when a widget is inserted or deleted, GVT mismatches the subsequent widgets if their positions are slightly shifted.
However, \tool proves to be more robust in correctly identifying the corresponding matches.
Second, GVT's strict matching threshold can overlook correct pairs.
In cases where widgets are swapped (Figure \ref{fig:gui-win} examples 5-6), GVT's strict matching threshold prevents it from identifying the swapping pair, as the score between them does not meet the threshold.
Instead, GVT reports these as individual missing widgets or extra widgets.
In contrast, \tool can correctly match the widgets with their swapping pairs, offering a more accurate report.

\paragraph{When \tool can have false positives?}

\begin{figure}[t]
    \centering
    \begin{subfigure}[t]{0.24\textwidth}
        \fbox{\includegraphics[width=\textwidth, keepaspectratio]{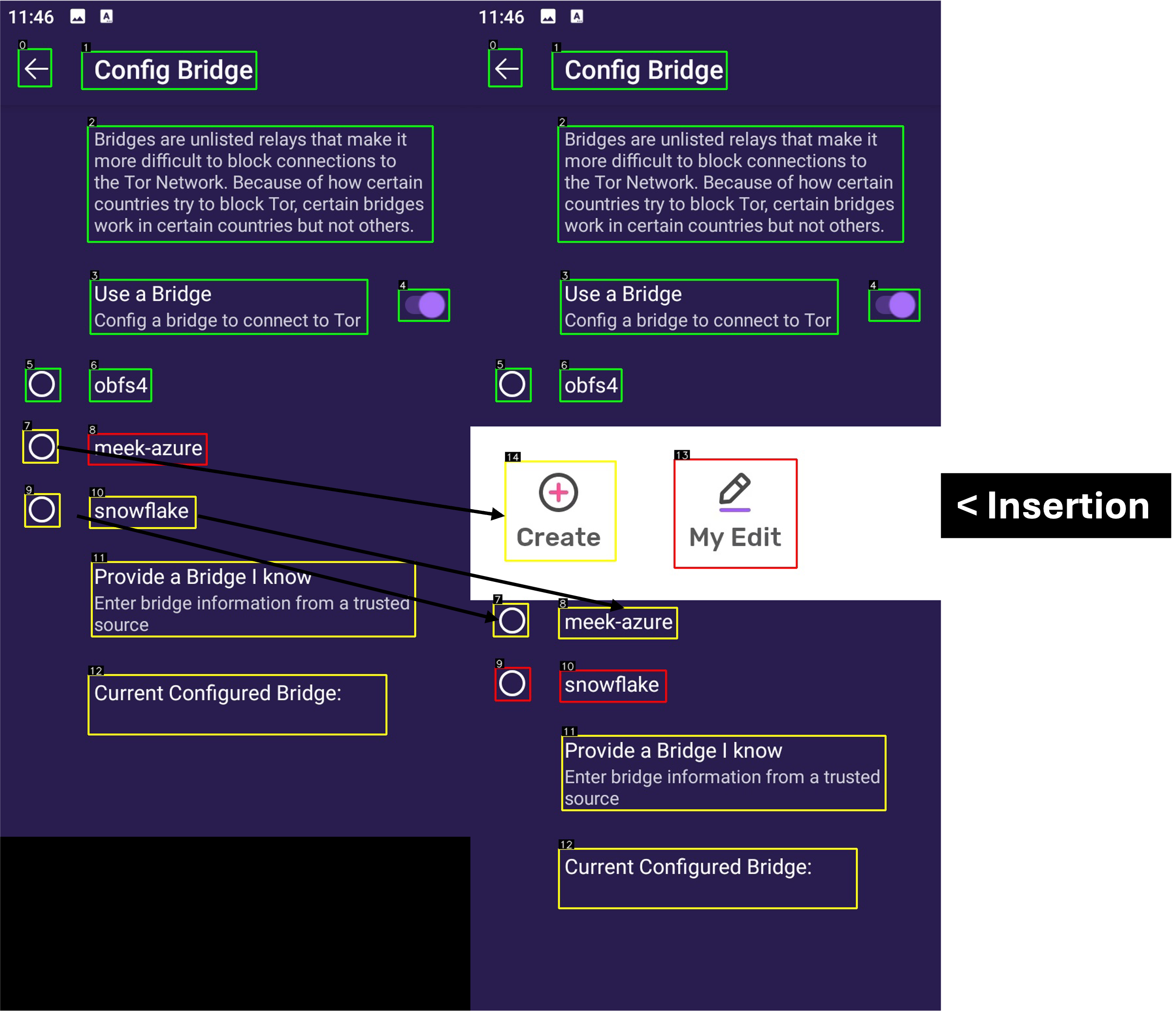}}
        \subcaption{FP Example 1}
    \end{subfigure}%
    \hfill
    \begin{subfigure}[t]{0.24\textwidth}
        \fbox{\includegraphics[width=\textwidth, keepaspectratio]{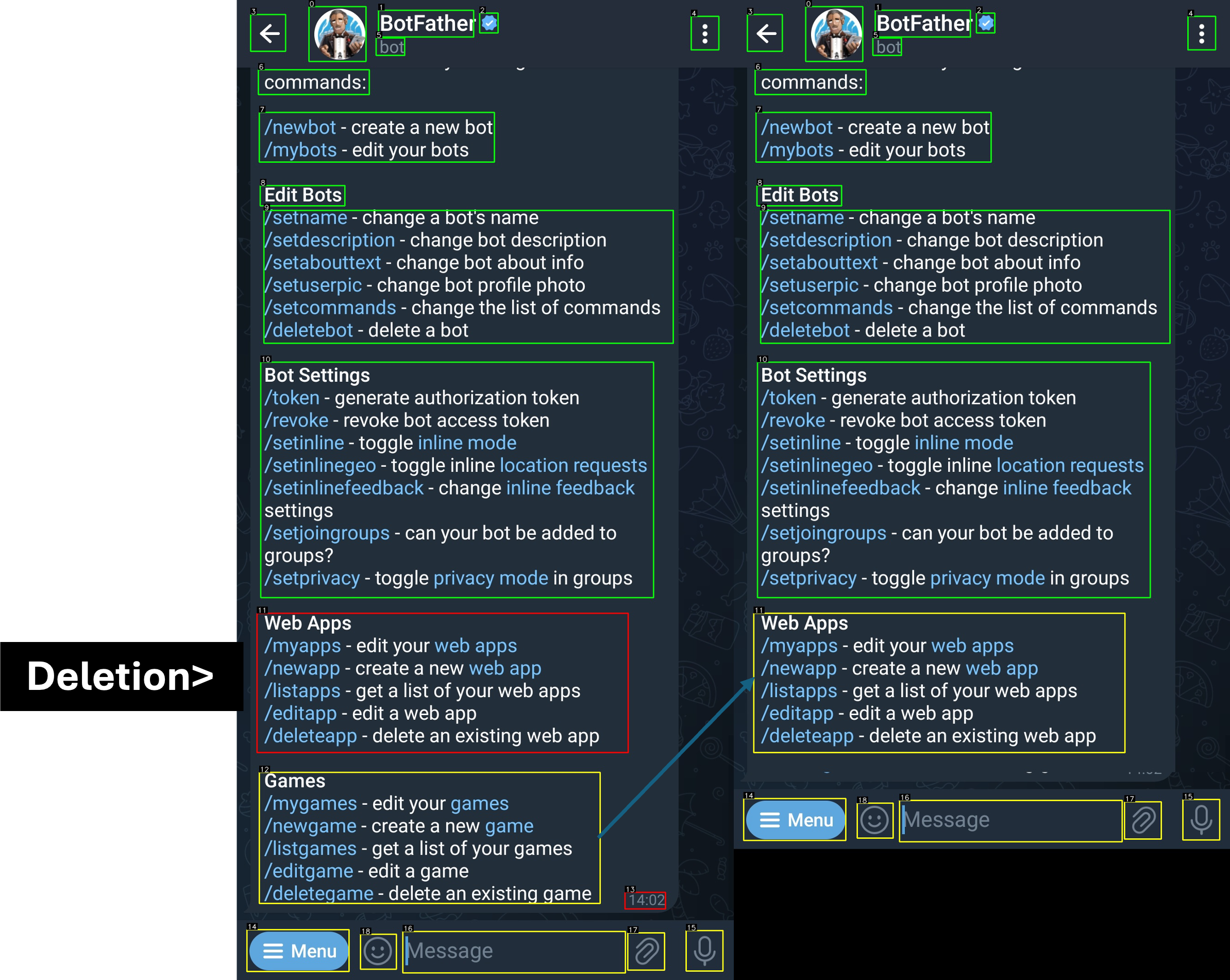}}
        \subcaption{FP Example 2}
    \end{subfigure}%
    \hfill
    \begin{subfigure}[t]{0.24\textwidth}
        \fbox{\includegraphics[width=\textwidth, keepaspectratio]{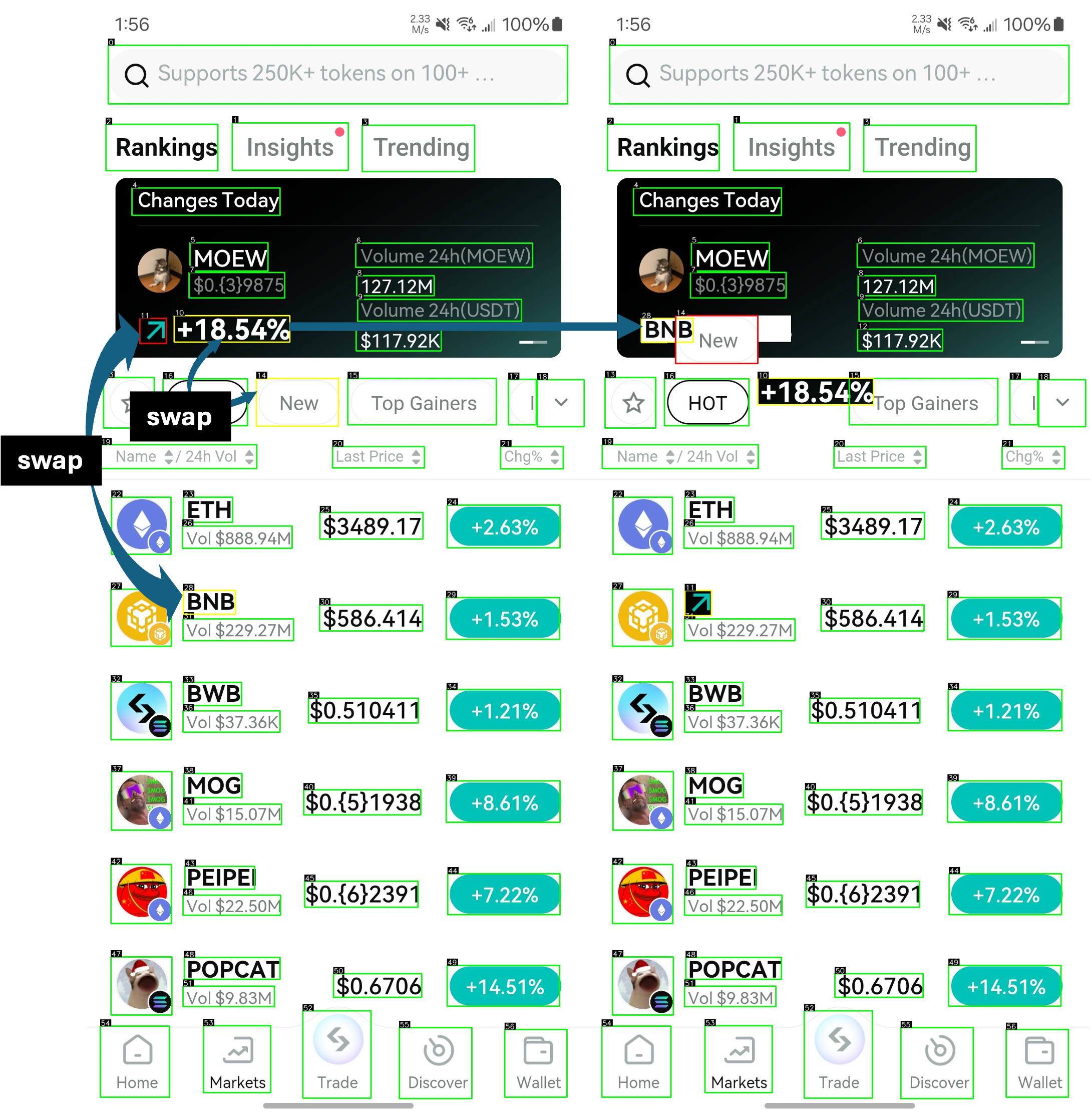}}
        \subcaption{FP Example 3}
    \end{subfigure}%
    \hfill
    \begin{subfigure}[t]{0.24\textwidth}
        \fbox{\includegraphics[width=\textwidth, keepaspectratio]{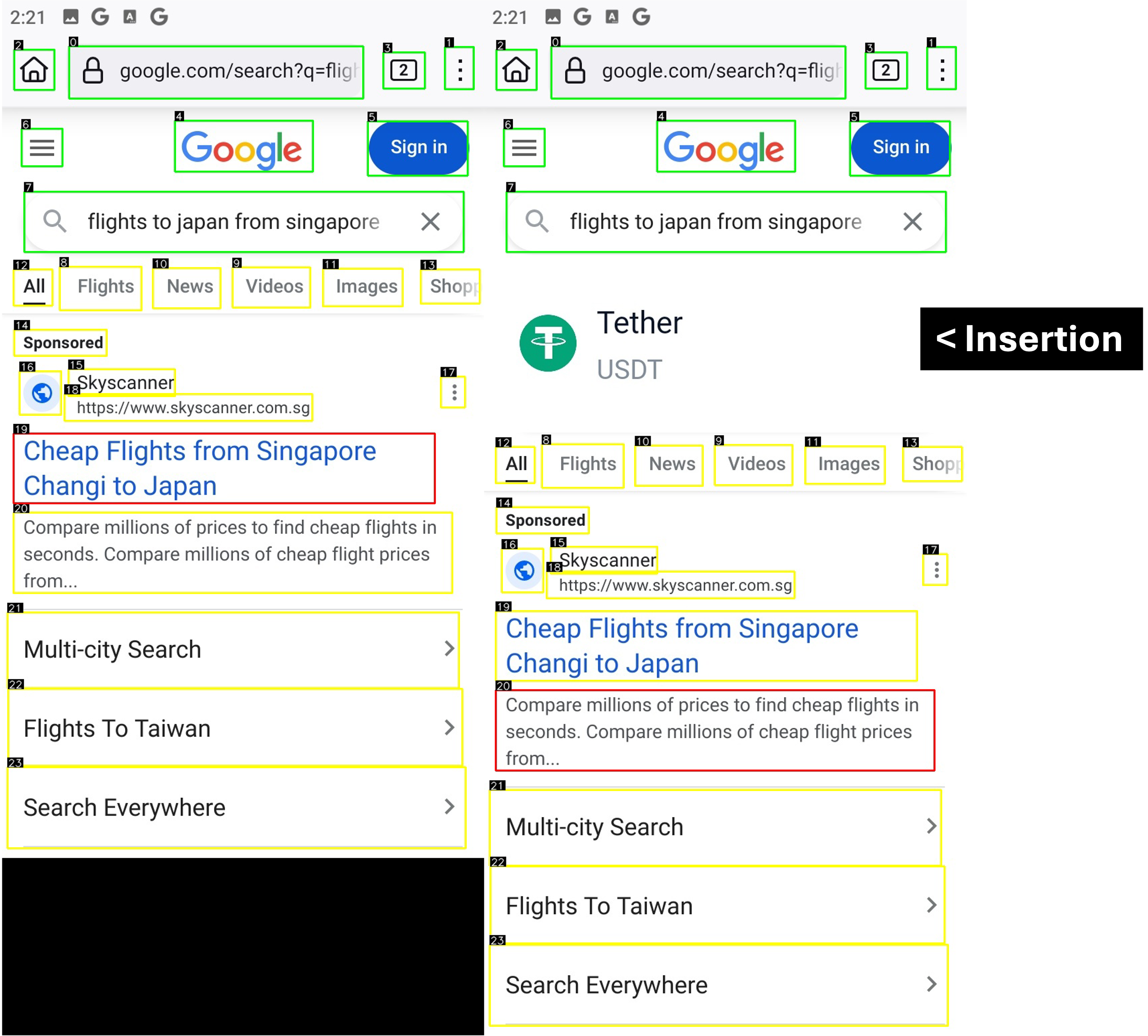}}
        \subcaption{FN Example 1}
    \end{subfigure}%
    \caption{
    False positive examples (FP) and false negative examples (FN) of \tool.
    In each figure, the original screen is displayed on the left, and the mutated screen (after insertion, deletion, or swapping) appears on the right.
    \coloredbox{SoftRed}{Red} boxes indicate extra or missing widgets.
    \coloredbox{SoftGreen}{Green} boxes denote widgets that are unaffected by the mutation.
    \coloredbox{SoftYellow}{Yellow} boxes those who have shifted due to the mutation but can still identify a match.
    \textbf{Lines} highlight certain matches reported by the approach.}

    \label{fig:gui-fail}
\end{figure}

We observe that \tool can have false positives when significant layout changes happen.
Although \tool is more resilient to layout perturbations than GVT, it can still produce mismatches when substantial changes occur, and the inserted or neighboring widget happens to share a similar type and shape.
As illustrated in Figure \ref{fig:gui-fail} (FP examples 1-2), an inserted or deleted widget may incorrectly pair with a neighboring one, causing subsequent widgets to become unpaired.
These unpaired widgets are then erroneously reported as missing or extra elements.
A potential remedy is to develop a more robust similarity metric that better captures the semantic and contextual similarities.
For example, we can train a metric learning model to compute similarity directly from widget appearances, ensuring accurate matching even under significant layout changes.

Further, we observe that overlaying widgets can also contribute to the false positive.
Widget swaps can occasionally lead to overlapping widgets (FP example 3).
This overlap may cause one widget to incorrectly match with a widget from another pair or interfere with the appearance, such as altering the color of the underlying widget.

\paragraph{When \tool can have false negatives?}
Similar to the reason for false positives, 
significant changes also contribute to false negatives.
When widgets are misaligned, actual missing or extra elements may be incorrectly paired with unrelated widgets, causing them to go unreported.
Further, we also observe that
the failure of the widget detector can also contribute to false negatives.
As shown in Figure \ref{fig:gui-fail} (FN example 1), the inserted widget was not detected by the object detector, resulting in a missed report.




\subsection{RQ2: Process Consistency Experiment}\label{exp:rq2}

\subsubsection{Setup}
We collect 160 design mock-ups across 80 applications.
However, we observed that some processes exhibited frequent version updates or blocked automated interactions,
thus, we retained 100 design mock-ups for the reproducibility of our experiment.

Previous studies \cite{xiong2023empirical, hu2024auitestagent, liu2024vision} have identified several common root causes of incorrect screen transitions in applications,
which include: 
(i) incorrect referencing of resource IDs, resulting in events being bound to the wrong buttons, 
(ii) lack of data synchronization, where global attributes—such as the user's login status—are not updated, and (iii) improper interception of events by parent or neighboring widget containers, causing touch events intended for a target widget to trigger the parent or neighboring widget instead.

Motivated by these observations, we randomly chose one screen transition represented by a <$\textbf{s}_{src}$, $desc$, $\textbf{s}_{tar}$> tuple for each of the 100 selected design mock-ups.
We then simulate incorrect screen transitions by randomly mutating the action in a selected screen transition to produce an incorrect target screen, thereby introducing process inconsistency.
Specifically, our mutation operation is chosen from one of the following:
\begin{enumerate}[leftmargin=*]
    \item \textbf{Target Mutation (Mutate $\textbf{s}_{tar}$).}
        The flow of the application returns to the previous screen instead of proceeding to the correct next screen. 
        This simulates situations where actions have no effect due to lack of data synchronization, causing the app to unexpectedly return to the previous screen.
        For example, when a user adds a new event to an empty calendar page, but after the event is created, the app returns to the empty calendar page instead of displaying the updated calendar.
    \item \textbf{Source Mutation (Mutate $\textbf{s}_{src}$)}. 
        The action is mutated to bind to a different widget.
        This simulates cases where the resource IDs are wrongly assigned due to similar functionality or mislabeling.
        For example, triggering the ``Log In'' event when the intended action was ``Sign Up''.
\end{enumerate}

We then validate those 100 mutated processes in the real application following the workflow in Figure \ref{app:process}.

We evaluate whether the execution halted at the mutated screen transition due to reported inconsistencies.
The outcomes were evaluated using precision and recall metrics, 
specifically assessing whether the screen matching correctly identified the true screen misalignments.
To ensure a stable and reproducible testing environment, 
we employed Waydroid \cite{waydroid} to deploy a virtual Android device on a Linux system.
All applications were installed on this device in their latest versions available during the experiments.
App interactions were automated using the UiAutomator2 driver \cite{uiautomator2} and the Android Debug Bridge (ADB).

\subsubsection{Results}
Our approach yielded Precision and Recall scores of 100\%, confirming that all introduced process inconsistencies were successfully detected.
The median runtime for each screen transition was 0.193 seconds.
The increase in runtime can be attributed to interaction delays inherent to in-app activities.

\subsection{RQ3-1: Widget Detection Performance}

\subsubsection{Setup}
In this experiment, we evaluate our trained widget detector's performance (Section \ref{app:widget-detection}).
We divide the collected screens (1392 screens) from public applications into training and testing datasets at a 7:3 ratio. 
The training set is used to train the widget detection model, specifically, the YOLO-v8 middle \cite{yolov8} object detection model. 
The testing set, on the other hand, serves to evaluate the model's generalization performance on previously unseen screens.
We use the standard metrics of mean Average Precision (mAP) and mean Average Recall (mAR) \cite{map} in object detection tasks. 
These metrics are computed at the Intersection-over-Union (IoU) threshold of 0.5. 
When the reported bounding box overlaps with the ground-truth box by more than the IoU threshold, it is recognized as a successful match.

\subsubsection{Results}
Table \ref{tab:widget-detection-accuracy} displays the class-wise performance of the widget detection model, which overall achieves a satisfactory detection rate (
As a reference, the state-of-the-art performance on COCO benchmark is around 0.502 \cite{ultralytics-yolov8}
). 
However, the accuracies for input boxes and charts are lower compared to other UI elements. 
The diminished accuracy for input boxes can be attributed primarily to their visual similarity to text buttons or combined buttons, which confuses the object detection model. 
Class confusion errors do not significantly impact our results. 
Even if an input box is misclassified as a text button on the mock-up screen, this misclassification would persist on the implementation screen. 
As a result, the two input boxes can still be aligned correctly.
To remedy this, implementing a post-check that verifies whether a widget supports text input actions based on its frontend code could be beneficial. 
Regarding charts, their detection is often compromised by the lack of clear boundaries that distinguish them from surrounding content, leading to failures in object detection.
Those failure examples can be found in \cite{rq3}.

\begin{table*}[t]
    \centering
    \caption{Widget detection accuracy.}\label{tab:widget-detection-accuracy}
    \resizebox{0.7\textwidth}{!}{%
    \begin{tabular}{llcc}
    \toprule
    Super-Category & Type & \textbf{mAP}@IoU=0.5 & \textbf{mAR}@IoU=0.5  \\
    \midrule
    \textbf{Interactable} & TextButton & 0.585 & 0.715 \\
    & IconButton & 0.711 & 0.804 \\
    & CombinedButton & 0.721 & 0.817\\
    & InputBox & 0.446 & 0.511 \\
    \midrule
    \textbf{Non-Interactable} & TextView & 0.677 & 0.758 \\
    & ImageView & 0.662 & 0.778 \\
    & Chart & 0.317 & 0.318 \\
    \midrule
    \textbf{Overall} & & 0.515 & 0.588 \\
    \bottomrule
    \end{tabular}
    }
\end{table*} 

\begin{figure}[h]
    \centering
    \includegraphics[width=0.5\linewidth]{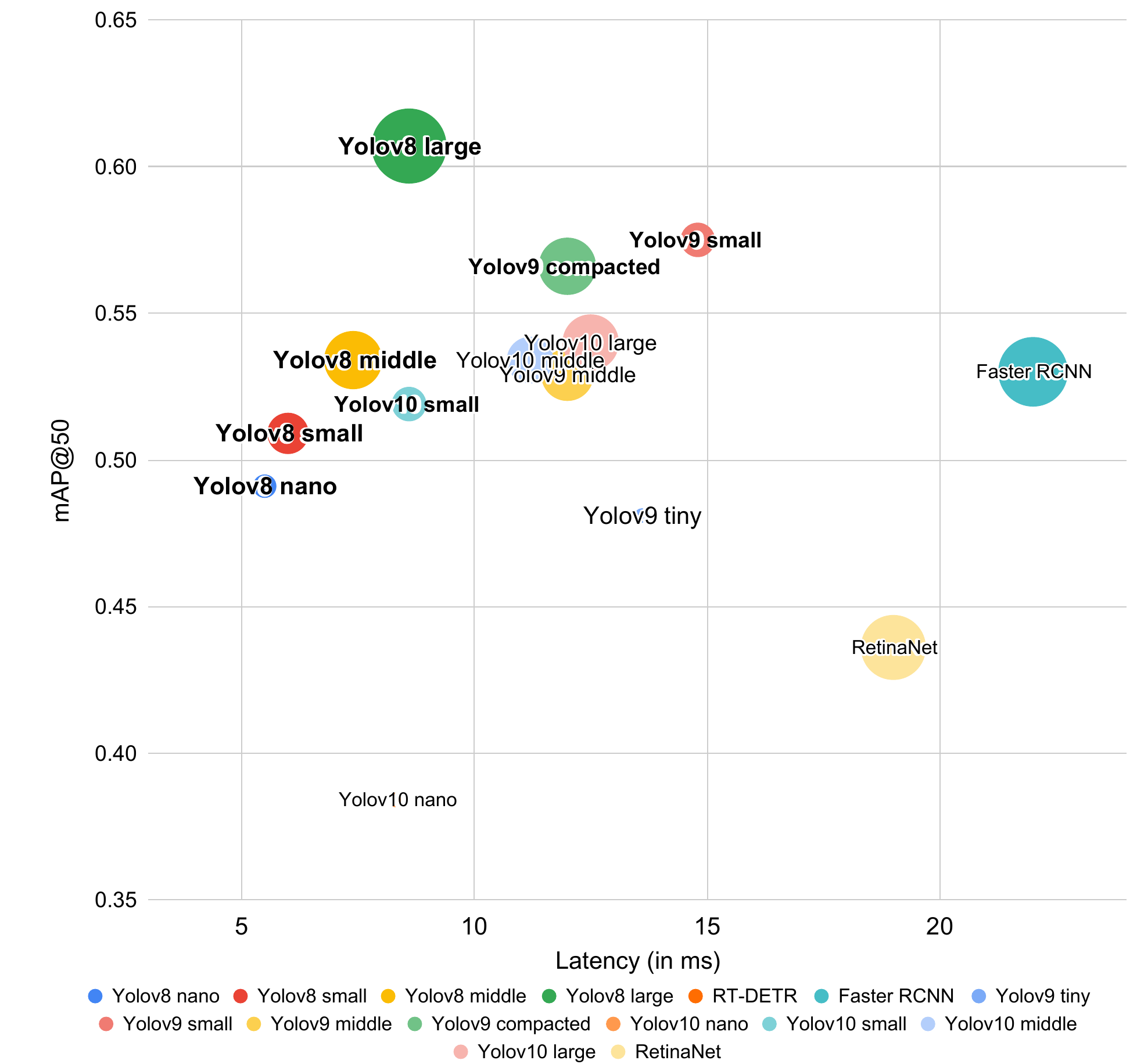}
    \caption{
    Comparison of Object Detection Architectures.
    The Y-axis represents the mean average precision, the X-axis represents the latency per image, and the bubble size corresponds to the parameter size.
    }
    \label{fig:object-detection-architect}
\end{figure}

In this work, we use the YOLO-v8 architecture.
We have also explored alternative detection methods such as other YOLO series \cite{yolov9, yolov10}, RetinaNet \cite{ross2017focal}, and two-stage Faster R-CNN model \cite{ren2016faster}.
\autoref{fig:object-detection-architect} shows the performance comparison.
From the comparison, we observe that the YOLO-v8 middle outperforms all other state-of-the-art models in terms of both precision and runtime efficiency while having a moderate parameter size.

\subsection{RQ3-2: VLM Action Completion Performance}

\subsubsection{Setup}
In this evaluation, we assess the performance of the VLM agent on 100 screen transitions (shared with RQ2) to determine whether it can accurately identify relevant widgets and convert actions into executable formats.
For each screen transition, the inputs to the VLM include natural language commands describing actions (e.g., ``click on the first suggestion'', ``expand menu'') and a GUI screenshot where interactable widgets are indexed and highlighted.

The VLM agent is expected to: (i) accurately pinpoint the widgets on the screen that correspond to the described actions, and (ii) effectively translate these natural language actions into executable commands.

We gauge the VLM's effectiveness through the transition success rate, which measures the agent's ability to correctly execute the specified actions and achieve the anticipated screen changes. 
For this experiment, we employ GPT-4o, the most advanced Vision-Language Model available at the time of submission, noted for its efficiency and cost-effectiveness in handling such tasks.

\subsubsection{Results}
Among the 100 transitions tested, 90 achieved the correct outcome within one trial, and 99 reached the correct outcome within two trials. 
This suggests that incorporating self-reflection loops could further enhance the accuracy of revisions. 
One specific transition failed because the pertinent widget was not detected by the object detector and, therefore, was not highlighted as an interactable widget when fed into the VLM, leading to its exclusion by the VLM.

\begin{figure}
    \centering
    \includegraphics[width=0.5\linewidth]{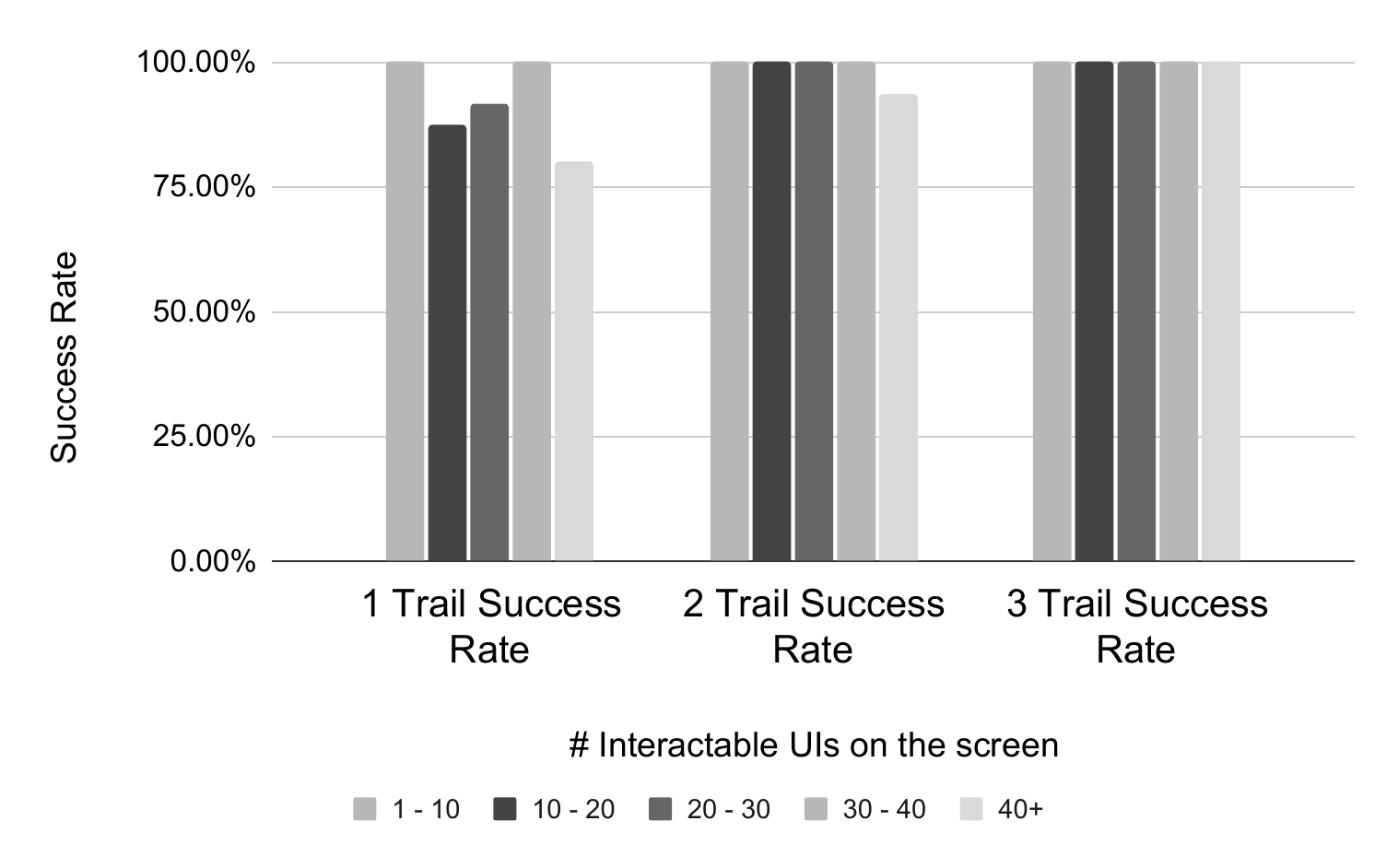}
    \caption{
    VLM Action Completion Performance by UI Layout
    }
    \label{fig:vlm-completion-grouped}
\end{figure}

We also examine the success rate across different UI layouts. 
Specifically, we categorize the 100 screen transitions based on the number of widgets present on the source screen: 1-10, 10-20, 20-30, 30-40, and 40+ widgets. 
The results are shown in \autoref{fig:vlm-completion-grouped}.
The darkness of the bars indicates the frequency of each UI layout, with the 10-20 widget range being the most common, followed by 20-30 widgets. 
The success rates are stable across different UI layouts.
There is no correlation between having more UIs and a lower success rate.

\section{Case Study}

In collaboration with a leading investment bank and financial services provider, we obtained the real app dataset.
This app is a mobile application designed to deliver comprehensive financial services to its users, including real-time stock trading, market analysis, financial news updates, portfolio management, and personalized investment advice.
We conduct a qualitative study on the trading app using real mock-ups.
In our collaboration with the company, we obtained 23 design mock-ups.
After discarding some outdated mock-ups, we focused on 19 design mock-ups.
We identified 8 instances of screen inconsistencies and 1 instance of process inconsistency.
Specifically, we observed 3 additional widgets, 9 missing widgets, 3 instances of inconsistent text widgets, and 5 instances of inconsistent color schemes in the implementation.
Apart from confirming the above inconsistencies in the application, 
the industrial experts send feedback that, 
the missing small widgets are particularly challenging for manual verification.
This underscores the essential role of automated testing tools in identifying and addressing such inconsistencies.
Those violations have been reported to and acknowledged by our industrial collaborators.

\begin{figure}
    \centering
    \begin{subfigure}[t]{0.32\textwidth}
        \fbox{\includegraphics[width=\textwidth, keepaspectratio]{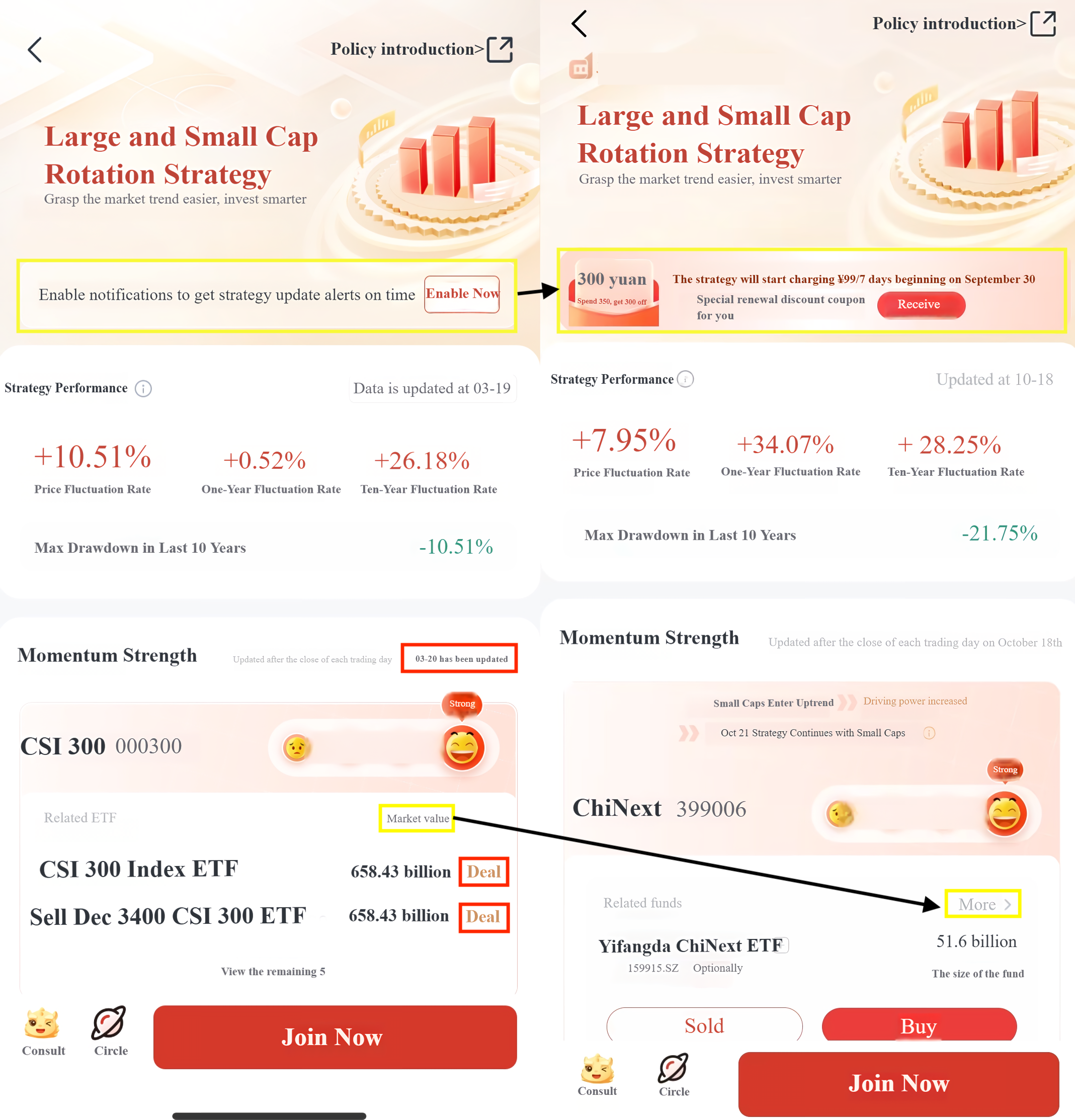}}
        \subcaption{Instance 1: semantic change and widget missing}
    \end{subfigure}%
    \hfill
    \begin{subfigure}[t]{0.32\textwidth}
        \fbox{\includegraphics[width=\textwidth, keepaspectratio]{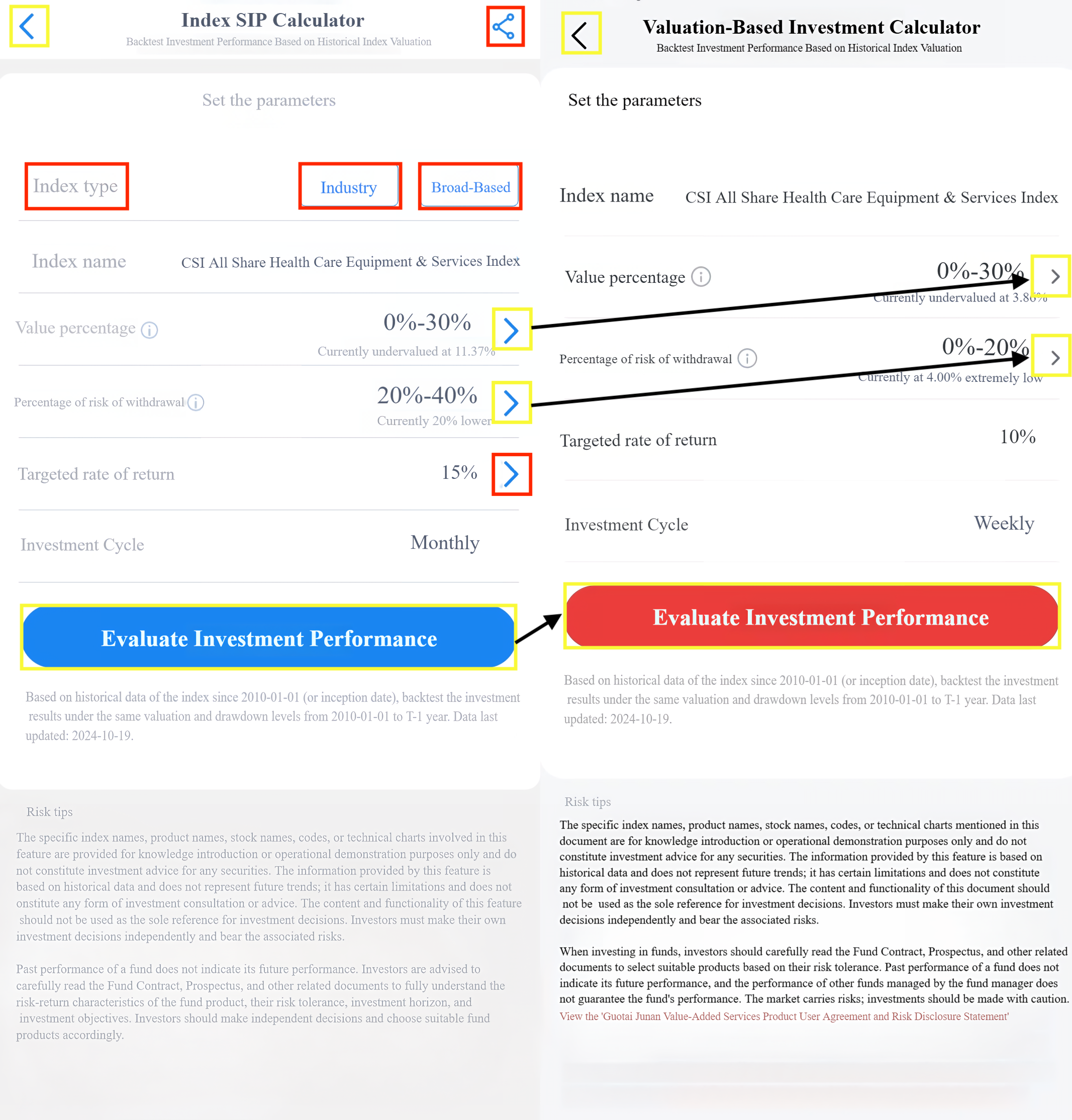}}
        \subcaption{Instance 2: semantic change and widget missing}
    \end{subfigure}%
    \hfill
    \begin{subfigure}[t]{0.32\textwidth}
        \fbox{\includegraphics[width=\textwidth, keepaspectratio]{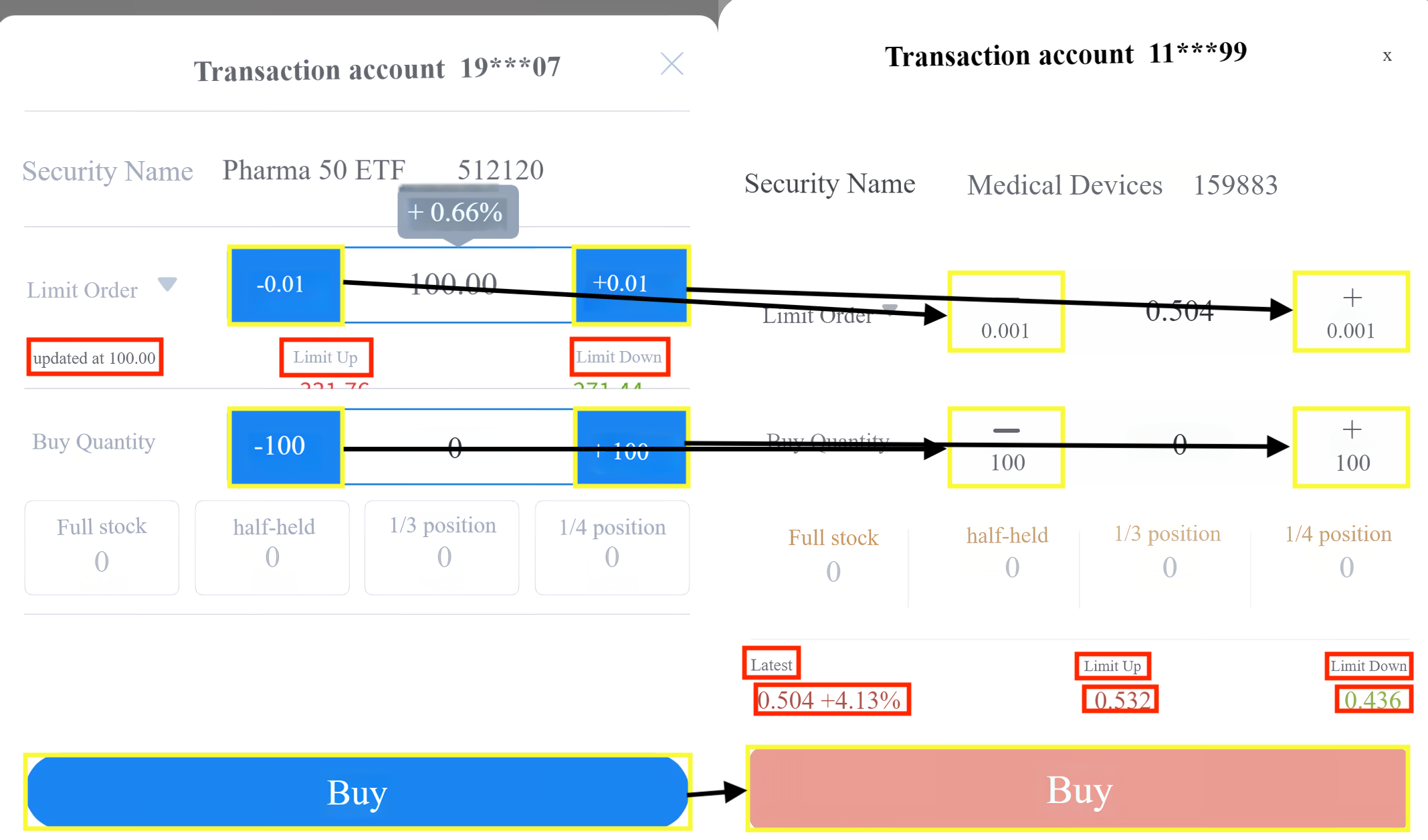}}
        \subcaption{Instance 3: semantic change, widget missing, and widget extra}
    \end{subfigure}%
    \caption{
    Real screen inconsistencies reported by the \tool on the trading App.
    In each figure, the mock-up is displayed on the left, and the implementation screen appears on the right.
    \coloredbox{SoftRed}{Red} boxes indicate extra or missing widgets.
    \coloredbox{SoftYellow}{Yellow} boxes indicate semantic inconsistencies.
    }
    \label{fig:case-study-screen-inconsistency}
\end{figure}

Figure \ref{fig:case-study-screen-inconsistency} shows instances of different types of inconsistencies.
In the first instance, the ``enable notification'' tab in the mock-up is replaced by a ``renew subscription'' tab in the implementation, the date that should be displayed in bold is not implemented, and the ``market value'' label is replaced with ``More >'' in the actual app.
In the second instance, the ``Index type'' label is missing, the sharing icon is absent, and several color discrepancies are observed between the mock-up and the final implementation.
More examples are shown on our anonymous website \cite{case-study}.


\subsection{Discussion}\label{sec:discussion}

\paragraph{Lack of Synchronization of Design Mock-ups}
In the real-world App development process, the design mock-up can be deprecated with respect to the current implementation.
This is because design mock-ups are not maintained with the same rigor as the mobile application.
In software development, code undergoes continuous integration and continuous deployment (CI/CD), ensuring it is regularly tested, updated, and refined.
Similarly, mock-ups require a systematic maintenance approach akin to CI/CD to ensure they evolve in parallel with the application to remain up-to-date.
Moreover, when developers implement changes to the application, they should communicate these modifications to the design team, ensuring the mock-ups are updated accordingly.
This calls for a strategic overhaul to enhance the efficacy and usability of mock-ups in development cycles.

\paragraph{Call for Good Practices in Mock-up Design}
Our case study on industrial mock-ups has revealed significant room for improvement in the current mock-up design approach.
Some important specification details are missing in the industrial design mock-ups.
In the study, it is non-trivial to navigate from the home screen in the application to the starting screen specified in the design mock-ups,
so that we can apply \tool.
In the study, we manually include such information in the industrial design mock-ups.

\paragraph{Threats to Validity}
(i) Internal Validity: Our tool depends on a third-party OpenAI service for action completion, so network latency and stability can impact performance. A future improvement is to distill a local large Visual-Language Model to alleviate these issues. 
(2) External Validity: 
In our simulation study, we use real mobile screenshots to simulate mock-ups due to the unavailability of actual design mock-ups. Our tool generalizes to other apps as long as (i) their mock-ups pass the compliance checks of the meta-model in Section \ref{app:meta-model} and (ii) each screen transition includes a description $desc$ (see Section \ref{app:process}). Any shortcomings can be manually corrected. 
\section{Related Work}
GUI testing has long been a focus of research, with studies aimed at improving efficiency, code coverage, fault localization \cite{lan2024deeply, yu2024practical, mahmud2024using, krishna2024motorease, bose2023columbus, ran2023badge, feng2023efficiency, ma2023automata, sun2023property, ahmed2023vialin, qian2022accelerating, su2022metamorphosis, alshayban2022accessitext, wang2022detecting, huang2021characterizing}, generating test cases \cite{zhang2024learning, talebipour2021ui, mariani2021semantic, hu2023omegatest, zhao2022avgust, song2017ehbdroid, saddler2017eventflowslicer, mirzaei2016reducing}, bug replay/reporting \cite{feng2024prompting, huang2024crashtranslator, huang2023context, feng2023read, feng2022gifdroid, yan2024semantic, bernal2022translating}, and code repair \cite{cao2024comprehensive, huang2023conffix, yang2023compatibility, zhang2023automated, zhao2022towards, alotaibi2021automated}.

\paragraph{Design Violation Detection in Mobile GUI} Vision-based approaches have been applied to detect GUI design violations. For instance, Seenomaly \cite{zhao2020seenomaly} and Nighthawk \cite{liu2022nighthawk} examine animations and layout issues using deep learning, while \cite{yang2021don} identifies common guideline breaches. These methods generally focus on generic guidelines, whereas GVT \cite{moran2018automated} specifically targets mock-up inconsistencies with a detailed taxonomy and a widget matching algorithm.

\paragraph{Mobile GUI Widget Detection and Matching} Early methods based on edge detection \cite{moran2018machine, nguyen2015reverse} and template matching \cite{bao2015scvripper, dixon2010prefab, qian2020roscript, yeh2009sikuli} have evolved into deep learning approaches for more accurate widget detection \cite{chen2019gallery, xie2020uied, chen2020object, ye2021empirical}. For widget matching, techniques in GVT \cite{moran2018automated}, METER, and MAPIT utilize layout distances, text edit thresholds, and graphic keypoints, although these pairwise criteria can be sensitive to perturbations \cite{mariani2021semantic}.

\paragraph{LLM-aided Mobile GUI Testing} Recent work leverages Large Language Models to improve mobile GUI testing. Systems like QTypist \cite{liu2023fill}, InputBlaster \cite{liu2024testing}, and HintDroid \cite{liu2024unblind} generate context-aware inputs, while GPTDroid \cite{liu2024make} and AutoDroid \cite{wen2024autodroid} employ LLMs for script generation and task automation.

\section{Conclusion}

This work introduces \tool, an innovative solution designed to detect inconsistencies between design mock-ups and actual application implementations.
\tool introduces the first comprehensive, end-to-end GUI testing solution that can detect both screen inconsistencies and process inconsistencies.
We tackled the screen-matching problem by proposing an accurate widget alignment optimization,
and the automatic process execution problem by introducing a visual prompt.
The application of \tool across various mobile applications has confirmed its utility and effectiveness.
Specifically, in a practical scenario involving a mobile app from a financial industry, \tool successfully identified real design violations, underscoring its value as a tool for improving the quality and reliability of software development through enhanced mock-up fidelity.

\section{Data Availability}
The datasets used in this study are publicly available and can be accessed via \cite{guipilot-website}.
The code used for data processing, analysis, and model training is available in the following repository \cite{code}. 

\section{Acknowledgment}
This research is funded by National Key Research and Development Program of China (Grant No.2023YFB4503802), 
the Minister of Education, Singapore (MOE-T2EP20124-0017), 
the National Research Foundation, Singapore, 
the Cyber Security Agency under its National Cybersecurity R\&D Programme (NCRP25-P04-TAICeN), 
DSO National Laboratories under the AI Singapore Programme (AISG Award No: AISG2-GC-2023-008), 
and CyberSG R\&D Cyber Research Programme Office. 

\newpage

\bibliographystyle{ACM-Reference-Format}
\bibliography{acmart}

\end{document}